\documentclass[12pt]{article}

\usepackage{amssymb}
\usepackage{amsmath}
\usepackage{latexsym}
\usepackage{yfonts}

\catcode`@=11
\renewcommand{\section}{\@startsection{section}{1}{0pt}{\medskipamount}
{\medskipamount}{\large\bf}}
\numberwithin{equation}{section}
\catcode`@=12

\def\beq{\begin{eqnarray}}    
\def\eeq{\end{eqnarray}}      

\def\pa{\partial}                       

\def\={\ =\ }

\begin{document}

\begin{center}

{\Large\bf Gauge dependence and multiplicative renormalization of  Yang-Mills
theory with matter fields}

\vspace{15mm}

{\large Igor A. Batalin$^{(a,b)}\footnote{E-mail:
batalin@lpi.ru}$\;,
Peter M. Lavrov$^{(b, c,d)} \footnote{E-mail:
lavrov@tspu.edu.ru}$,\;
Igor V. Tyutin$^{(a,b)}\footnote{E-mail:
tyutin@lpi.ru}$\;
}

\vspace{8mm}

\noindent ${{}^{(a)}}$
{\em P.N. Lebedev Physical Institute,\\
Leninsky Prospect \ 53, 119 991 Moscow, Russia}

\noindent  ${{}^{(b)}}
${\em
Tomsk State Pedagogical University,\\
Kievskaya St.\ 60, 634061 Tomsk, Russia}

\noindent  ${{}^{(c)}}
${\em
National Research Tomsk State  University,\\
Lenin Av.\ 36, 634050 Tomsk, Russia}

\noindent  ${{}^{(d)}}
${\em
Departamento de F\'isica, ICE,\\ Universidade Federal de Juiz de Fora,\\
Campus Universit\'ario-Juiz de Fora,\\ 36036-900, MG, Brazil}

\vspace{15mm}

\begin{abstract}
\noindent In the paper, within the background field method, the
renormalization and the gauge dependence is studied as for an SU(2)
Yang-Mills theory  with  multiplets of spinor and scalar fields. By
extending the quantum action of the BV-formalism  with an extra
fermion vector field and a constant fermion parameter, the
multiplicative character of the renormalizability is proven. The
renormalization of all the physical parameters of the theory under
consideration is shown to be gauge-independent.

\vfill
\end{abstract}

\end{center}

\vfill

\\

\newpage

\section{Introduction}
\noindent When constructing modern models of fundamental
interactions \cite{Weinberg}, non-Abelian gauge field theories
\cite{YM}  play a central role. Any gauge theory can be quantized in
a covariant way within the BV-formalism \cite{BV,BV1} involving the
gauge-fixing procedure as an important tool. Although at the quantum
level the gauge symmetry is broken, nevertheless, it causes the
existence of fundamental global supersymmetry known as the BRST
symmetry \cite{BRS1,T}; application of the BRST symmetry  to
renormalization of gauge theories has been proposed in paper \cite{BRS2}.
The respective conserved fermion nilpotent
generator is known as the BRST charge  \cite{BV,BV1} responsible for
correct construction of physical state space \cite{KO, Henneaux}.
Due to the equivalence theorem \cite{KT} and the BRST symmetry, it
is succeeded to prove that the physical $S$-matrix is independent of
the choice of gauge fixing.

 The gauge dependence problem did appear by itself from the
study of the effective potential,  which appeared to be
gauge-dependent in Yang-Mills theories with the spontaneous symmetry
breaking, when calculating physically-sensible results (the energy
of the ground state, the masses of the physical particles, and so
on) \cite{Jac, DJac}. In Refs. \cite{Niel,FK} it was established
that the energy of the ground state was gauge-independent. Later, it
was proved \cite{LT3,LT1} that in Yang -Mills theories the
dependence of gauge parameters in the effective action could be
described in terms of gauge-invariant functional whose arguments
(fields) were gauge-dependent (see also recent Refs. \cite{Niel1,PT}
devoted to that problem as resolved via the procedure of
redefinition of the field variables, found in \cite{LT3,LT1}).
When studying the gauge dependence problem it was a nice idea as to enlarge
the usual BRST transformations by transforming also the gauge parameter $\xi$
into fermion variable $\chi$, which results in the extension of the master equation
\cite{PS}. We have used a similar extension in the present paper, together with
introducing new fermion fields $\theta^a_{\mu}$  \cite{K-SZ}.

Our investigation  of renormalization and gauge dependence in an
SU(2) Yang-Mills theory with spinor and scalar fields is based on
using the background field formalism \cite{DeW,AFS,Abbott}. This
formalism is  the popular method for quantum studies  and
calculations in gauge theories because it allows one to work with
the effective action invariant under the gauge transformations of
the background fields, and to reproduce all usual physical results
which can be obtained within the standard quantization approach.
Various aspects of quantum properties of Yang-Mills theories and
quantum gravity theories have been successfully studied in this
technique \cite{'tH,K-SZ,GvanNW,CMacL,IO,GS,Ven,Gr} (among recent
applications see, for example,  \cite{Barv,FT,BLT-YM,BFMc}).

Although there are many papers devoted to various aspects of renormalizability
of Yang-Mills theories, the gauge dependence of the renormalization constants
has been studied explicitly only as for the gauge field sector \cite{K-SZ}. In
recent paper \cite{BLT-YM}, we have studied a structure of the
renormalization procedure and a gauge dependence as for an Yang-Mills theory
with a multiplet of spinor fields based on an arbitrary simple compact gauge
Lie group. We have proved the multiplicativity of the renormalization and the
gauge independence of the renormalization constants of the physical
parameters of the theory (gauge interaction constant and fermion mass
parameter). The main goal of the present paper is to include scalar fields
into a gauge model as to have the possibility of generating masses to physical
particles through the spontaneous symmetry breaking \cite{Higgs}. Thereby, we
have generalized the previous results to the maximal spectrum of spins,
$\{0,1/2,1\}$, capable to meet the requirement of multiplicative
renormalizability. In contrast to the case of spinor fields, the multiplicative
renormalizability in the sector of scalar fields is not provided with the gauge
symmetry only, and this requires for an additional proof. Also, specific
problems may appear when studying the gauge independence of
renormalization  constants of new couplings of scalar fields which do
not connect directly with Yang-Mills interactions. Thus, the full description
of the renormalization structure requires for a detailed explicit analysis of
the respective coupled equations, which is presented in the paper. To simplify
the presentation and calculations, we restrict ourselves by the
simplest Yang-Mills theory with SU(2) gauge group. Generalization of our results
to another gauge group can be made straightly.

The paper is organized as follows. In Section 2,  the action of
SU(2) Yang-Mills theory with multiplets of  spinor and scalar fields
in the background field method is extended with the help of
additional fermion vector field and constant fermion parameter which
allows one later to arrive at a  multiplicative renormalizable
theory. The symmetries of this extended action are studied and
presented
 in a set of equations for that action.
In Section 3,  it is established the structure and the arbitrariness
is described  for any local functional with the quantum numbers of
the extended action that satisfies the same set of equations as the
extended action does. In Section 4, the equations are derived for
the generating functional of vertexes (effective action), as a
consequence at the quantum level, of the symmetry property of the
extended action; and it is shown that the generating functional of
vertexes satisfies the same equations as the extended quantum action
does. In Section 5, it is studied the renormalization procedure of
the theory considered when using the loop expansion technique and
the minimal subtraction scheme; and thus the multiplicative
renormalizability of the theory is proved. In Section 6, the gauge
independence of all physical parameters of the theory under
consideration to any order of loop expansions is found. Concluding
remarks are given  in Section 7.

Condensed DeWitt's notations \cite{DeWitt} are used through the paper.
Functional derivatives with
respect to field variables are understood as the left. Right derivatives
of a quantity $f$ with respect to the variable $\varphi$
are denoted as $f\frac{\overleftarrow{\delta}}{\delta \varphi}$.

\section{Extended quantum action}
\noindent
Let us consider an $SU(2)$-gauge theory of non-Abelian
vector fields $A^{\alpha}_{\mu}=A^{\alpha}_{\mu}(x)$, a doublet of
spinor fields $\psi_j=\psi_j\;\!(x),
\overline{\psi}_j=\overline{\psi}_j\;\!(x)$ and a triplet of real
scalar fields  $\varphi^\alpha=\varphi^\alpha(x)$ in the d=4
Minkowski space-time, with the action
\beq
\label{a1}
&&\mathcal{S}_{YM}(A,\Psi,\varphi)=\mathcal{S}_1(A)+\mathcal{S}_2(A,\Psi)+
\mathcal{S}_3(A,\varphi)+\mathcal{S}_4(\Psi)+\mathcal{S}_5(\varphi)+
\mathcal{S}_6(\Psi,\varphi), \\
\label{a1a}
&&\mathcal{S}_1(A)=-\frac{1}{4}\int dx G_{\mu\nu}^\alpha(A)G_{\mu\nu}^
\alpha(A),\quad \mathcal{S}_2(A,\Psi)=\int dx i\overline{\psi}_j\gamma^\mu
D_{\psi\mu jk}(A)\psi_k, \\
\label{a1b}
&&\mathcal{S}_3(A,\varphi)=\frac{1}{2}\int dx D_{\varphi\mu}^{\alpha\beta}(A)
\varphi^\beta D_{\varphi\mu}^{\alpha\gamma}(A)\varphi^\gamma, \quad
\mathcal{S}_4(\Psi)=-m\int dx \overline{\psi}_j\psi_j, \\
\label{a1c}
&&\mathcal{S}_5(\varphi)=\int dx\Big(-\frac{M^2}{2}\varphi^2-\frac{\lambda}{4}
\varphi^4\Big), \quad \mathcal{S}_6(\Psi,\varphi)=i\vartheta\int dx
\varphi^\alpha\overline{\psi}_jt_{jk}^\alpha\psi_k,
\eeq
where the notations
\beq
\nonumber
&&G_{\mu \nu }^{\alpha }(A)=\pa_{\mu}A^{\alpha}_{\nu}-
\pa_{\nu}A^\alpha_\mu +g\varepsilon^{\alpha\beta\gamma}A_\mu^\beta
A_\nu^\gamma, \\
\nonumber
&&D_{\psi\mu jk}(A)=\delta_{jk}\pa_{\mu}+gt^\alpha_{jk}A^\alpha_\mu, \quad
t^\alpha_{jk}=\frac{i}{2}\sigma^\alpha_{jk}, \quad
\Psi=\{\psi,\overline{\psi}\}, \\
\label{a2}
&&D_{\varphi\mu}^{\alpha\beta}(A)=\delta_{\alpha\beta}\pa_\mu+
gE_{\alpha\beta}^\gamma A_\mu^\gamma, \quad \varphi^2=
\varphi^\alpha\varphi^\alpha, \quad \varphi^4=(\varphi^2)^2, \\
\nonumber &&E_{\alpha\beta}^\gamma=\varepsilon^{\alpha\gamma\beta},
\quad \alpha,\gamma,\beta=1,2,3, \quad j,k=1,2,
\eeq
were used. In
the relations  (\ref{a1}) - (\ref{a1c}) and (\ref{a2})
$\varepsilon^{\alpha\beta\gamma}$ represent the structure
coefficients of the  $SU(2)$ gauge group,
$t^\alpha=\{t^\alpha_{jk}\}$ and
$E^{\alpha}=\{E^{\alpha}_{\beta\gamma}\}$ are the generators of
gauge transformations of spinor and scalar fields with the
properties \beq \label{a3} [E^\alpha,E^\beta]=
\varepsilon^{\alpha\beta\gamma}E^\gamma, \quad [t^\alpha,t^\beta]=
\varepsilon^{\alpha\beta\gamma}t^\gamma,\quad
(t^{\gamma})^+=-t^\gamma, \quad[\gamma^{\mu},t^{\alpha}]=0, \eeq
$\gamma^{\mu}$ are the Dirac matrices, $\sigma^\alpha_{jk}$ are the
Pauli matrices and $g$ is a gauge coupling parameter, $\lambda$ is a
coupling constant of the scalar field, $\vartheta$ is a coupling
constant of the scalar and spinor fields,  $m$ and $M$ are mass
parameters of the spinor and the scalar fields, respectively.

The action  (\ref{a1}) is invariant under the
$SU(2)$-gauge transformations with gauge parameters
$\omega_{\alpha}=\omega_{\alpha}(x)$,
\beq
\nonumber
&&\delta_{\omega}\mathcal{S}_{YM}(A,\Psi,\varphi)=0 \\
\nonumber
&&\delta_{\omega}A_\mu^\alpha= \left(\delta_{\alpha\beta}\partial_\mu+
g\varepsilon^{\alpha\sigma\beta}A_\mu^\sigma\right)\omega _{\beta }=
D_\mu^{\alpha\beta}(A)\omega_\beta, \\
\nonumber
&&\delta _\omega\psi_j=-gt_{jk}^\beta\psi_k\omega_\beta, \quad
\delta_\omega\overline{\psi}_j=g\overline{\psi}_kt_{kj}^\beta\;\!\omega_\beta,
 \\
\label{a4}
&&\delta_\omega\varphi^\alpha=-
gE_{\alpha\gamma}^\beta\varphi^\gamma\omega_\beta.
\eeq

Notice that the form  (\ref{a1c}) of the polynomial  $\mathcal{S}_5(\varphi)$
is uniquely determined by the invariance requirement  under global
$SU(2)$-transformations.

In the background-field formalism  \cite{DeW,AFS,Abbott} the gauge
field $A^\alpha_\mu$ appearing in the classical action (\ref{a1}) is
replaced by $A^\alpha_\mu+{\cal B}^\alpha_\mu$,
\beq
\label{a7}
\mathcal{S}_{YM}(A,\Psi,\varphi)\;\rightarrow \;
\mathcal{S}_{YM}(A+{\cal B},\Psi,\varphi),
\eeq
where ${\cal
B}^{\alpha}_{\mu}$ is considered as an external vector field.
Effective action for functional integral in quantum theory is
constructed with using the BV-formalism \cite{BV,BV1}.

To study a renormalization structure and a gauge dependence of
renormalization constants we use an extended action. This action
$S_{ext}=S_{ext}(Q,Q^*,\overline{C},B,{\cal B},\xi,\theta,\chi)$ is
constructed by introducing additional fermion fields
$\theta^\alpha_\mu=\theta^\alpha_\mu(x)$ and  a constant fermion
parameter\footnote{For the first time such additional variables were
used in  \cite{K-SZ}.} $\chi$,  and has the form
\beq
\nonumber
&&S_{ext}=\mathcal{S}_{YM}(A+{\cal B},\Psi,\varphi)+\int dx
Q^{*}{\cal R}_Q+ \int
dx\Big(\overline{C}^{\alpha}D^{\alpha\gamma}_\mu({\cal B})
D^{\gamma\beta}_\mu(A+{\cal B})C^\beta + \\
\nonumber
&&\qquad+B^\alpha D^{\alpha\beta}_\mu({\cal B})A^\beta_\mu+
(\xi/ 2)B^\alpha B^\alpha\Big)+ \int dx\Big(\theta^\alpha_\mu
[D^{\alpha\beta}_\mu(A+\mathcal{B})\overline{C}^\beta-
A^{\ast\alpha}_\mu]+ \\
\label{a11}
&&\qquad+\chi[(A^{\ast\alpha}_\mu-
D^{\alpha\beta}_\mu(\mathcal{B})\overline{C}^\beta)A^{\alpha}_\mu
+C^{\ast\alpha}C^{\alpha}+\psi^\ast_j\psi_j+\overline{\psi}^\ast_j
\overline{\psi}_j+\varphi^{\ast\alpha}\varphi^\alpha]\Big),
\eeq
where $Q$  means the set of fields  $\{A^\alpha_\mu,\psi_j,\overline{\psi}_j,
\varphi^{\alpha},C^\alpha\}$, the symbol $Q^*$ is used for the set of
corresponding  antifields and ${\cal R}_Q$ are generators of the BRST
transformations \cite{BRS1,T},
\beq
\nonumber
&&{\cal R}_{A^\alpha_\mu}=D^{\alpha\beta}_\mu(A+{\cal B})C^\beta, \quad
{\cal R}_{\psi_j}=-gt^\alpha_{jk}\psi_kC^\alpha,\quad
{\cal R}_{\overline{\psi}_j}=g\overline{\psi}_kt^\alpha_{kj}C^\alpha, \\
\label{a6}
&&{\cal R}_{\varphi^\alpha}=-gE_{\alpha\gamma}^\beta\varphi^\gamma C^\beta,
\quad {\cal R}_{C^\alpha}=\frac{g}{2}\varepsilon^{\alpha\beta\gamma}
C^\beta C^\gamma.
\eeq

The action  (\ref{a11}) is invariant,
\[
S_{ext}\frac{\overleftarrow{\delta}}{\delta\Pi_I}\delta\Pi_I\epsilon=0,
\]
under the following global supersymmetry transformations of the
variables (here $\{\Pi_I\}$ is the set of all the variables entering
the action, $\epsilon$ is a constant fermion parameter of the
transformation, $\varepsilon(\epsilon)=1$, $\epsilon^2=0$)),
\beq
\label{a13} &&\delta A^{\alpha}_{\mu}=
D^{\alpha\beta}_{\mu}(A+\mathcal{B})C^{\beta}+\theta^{\alpha}_{\mu}=
\frac{\delta }{\delta A^{\ast \alpha}_{\mu}}S_{\mathrm{ext}}+
\chi A^{\alpha}_{\mu}, \\
\nonumber
&&\delta B^{\alpha}=-\frac{1}{\xi }\big[D^{\alpha\gamma}_{\mu}(\mathcal{B})
D^{\gamma\beta}_{\mu}(A+\mathcal{B})C^{\beta}+D^{\alpha\beta}_{\mu}(A+
\mathcal{B})\theta^{\beta}_{\mu} + \chi
D^{\alpha\beta}_{\mu}(\mathcal{B})A^{\beta}_{\mu}\big]-\chi B^{\alpha}= \\
\label{a14}
&&=-\frac{1}{\xi }\frac{\delta }{\delta \overline{C}^{\alpha}}
S_{\mathrm{ext}}-\chi B^{\alpha}, \\
\label{a15}
&&\delta C^\alpha=\frac{g}{2}\varepsilon^{\alpha\beta\gamma}C^{\beta}
C^{\gamma }=\frac{\delta }{\delta C^{*\alpha}}S_{\mathrm{ext}}-\chi C^{\alpha }, \\
\label{a16}
&&\delta \overline{C}^{\alpha }=
-\frac{1}{\xi }D^{\alpha\beta}_{\mu}(\mathcal{B})A^{\beta}_{\mu}+
\chi \overline{C}^{\alpha }=-\frac{1}{\xi }\frac{\delta }{\delta B^{\alpha}}
S_{\mathrm{ext}}+B^{\alpha }+\chi \overline{C}^{\alpha }, \\
\label{a17}
&&\delta \psi_j =-gt^{\alpha }_{jk}\psi_k C^{\alpha }=
\frac{\delta }{\delta \psi
^{\ast }_j}S_{\mathrm{ext}}-\chi \psi_j ,\\
\label{a18}
&&\delta \overline{\psi }_j=g\overline{\psi}_kt^{\alpha}_{kj}C^{\alpha}=
\frac{\delta}{\delta\overline{\psi}^{\ast}_j}S_{\mathrm{ext}}-
\chi\overline{\psi }_j, \\
\label{a19}
&&\delta\varphi^\alpha=-gE_{\alpha\gamma}^\beta \varphi^\gamma C^\beta=
\frac{\delta}{\delta\varphi^{\ast\alpha}}S_{\mathrm{ext}}+
\chi\varphi^\alpha, \\
\label{a20}
&&\delta A^{\ast \alpha}_{\mu}=\chi A^{\ast \alpha}_{\mu},\;
\delta C^{\ast \alpha}=-\chi C^{\ast \alpha},\quad\!\!
\delta\psi^\ast_j=-\chi\psi^\ast_j,\quad\!\! \delta\overline{\psi}^\ast_j=
-\chi\overline{\psi}^\ast_j, \quad\!\! \delta\varphi^{\ast\alpha}=
\chi\varphi^{\ast\alpha}, \\
\label{a21} &&\delta \xi =2\xi \chi ,\quad \delta
\mathcal{B}^{\alpha}_{\mu}=- \theta^{\alpha}_{\mu} , \quad \delta
\theta^{\alpha}_{\mu}=0,\quad \delta \chi =0.
\eeq
Taking into
account the right-hand sides of the relations   (\ref{a13}) -
(\ref{a16}) and omitting indices of all variables,  the invariance
condition of the action $S_{ext}$ rewrites  in the form of the
following equation
\beq
&&\int
dx\Big(S_{\mathrm{ext}}\frac{\overleftarrow{\delta }}{\delta Q}
\frac{\delta}{\delta Q^\ast}S_{\mathrm{ext}}-B\frac{\delta}{\delta
\overline{C}}S_{\mathrm{ext}}-\theta \frac{\delta }
{\delta \mathcal{B}}S_{\mathrm{ext}}\Big) +  \notag \\
\label{a22}
&&+\chi\int dx\Big[Q\frac{\delta}{\delta Q}-Q^*\frac{\delta}{\delta Q^*}
-\overline{C}\frac{\delta}{\delta\overline{C}}-B\frac{\delta}{\delta B}
\Big]S_{\mathrm{ext}}+2\chi\xi\frac{\partial}{\partial\xi}S_{\mathrm{ext}}=0.
\eeq

Also, the action (\ref{a11}) satisfies the equation
\beq
\label{a23}
S_{\mathrm{ext}}\overleftarrow{H^\alpha}\omega_{\alpha }=0,
\eeq
where
\beq
\label{a23a} &&\overleftarrow{H^\alpha}\omega_\alpha=\int
dx\Big\{\Big
[\frac{\overleftarrow{\delta}}{\delta\mathcal{B}_\mu^\beta}D_\mu^{\beta\alpha}
(\mathcal{B})+g\varepsilon^{\beta
\gamma\alpha}\Big(\frac{\overleftarrow {\delta}}{\delta
A_\mu^\beta}A_\mu^\gamma+\frac{\overleftarrow{\delta}} {\delta
B^\beta}B^\gamma+\frac{\overleftarrow{\delta}}{\delta
C^\beta}C^\gamma+ \frac{\overleftarrow{\delta}}
{\delta\overline{C}^\beta}\overline{C}^\gamma\Big)- \notag \\
&&-gE_{\beta\gamma}^\alpha\Big(\frac{\overleftarrow{\delta}}
{\delta\varphi^\beta}\varphi^\gamma+\frac{\overleftarrow{\delta}}
{\delta\varphi^{*\beta}}\varphi^{*\gamma}\Big) -
gt_{jk}^\alpha\Big(\frac{\overleftarrow{\delta}}{\delta\psi_j}\psi_k+
\frac{\overleftarrow{\delta}}{\delta\overline{\psi}_j^\ast}
\overline{\psi}_k^{\;\!\ast}\Big)+g\Big(\frac{\overleftarrow{\delta}}
{\delta\overline{\psi}_j}\overline{\psi}_k+\frac{\overleftarrow{\delta}}
{\delta\psi_j^\ast}\psi_k^\ast\Big)t_{kj}^\alpha+ \notag \\
&&+g\varepsilon^{\beta\gamma\alpha}\Big(\frac{\overleftarrow{\delta}}
{\delta
A_\mu^{\ast\beta}}A_\mu^{\ast\gamma}+\frac{\overleftarrow{\delta}}
{\delta C^{\ast \beta }}C^{\ast \gamma }+\frac{\overleftarrow{\delta
}} {\delta\theta_\mu^\beta}\theta
_\mu^\gamma\Big)\Big]\omega_\alpha\Big\},
\eeq
is the operator of
the gauge transformations in the sector of the fields
$\mathcal{B}_{\mu }$, $\varphi$, $\psi$, $\overline{\psi}$ and, at
the same time, of the tensor transformations in the sector of the
fields $A_{\mu }$, $C$, $\overline{C}$, $B$, $\theta_\mu $,
$A_\mu^*$, $\varphi^*$, $\psi^*$, $\overline{\psi}^\ast$, $C^\ast$.

Finally, let us notice the two important relations linear in the
fields $A_\mu,B$ and their derivatives, which the action (\ref{a11})
satisfies to,
\beq
\label{a24}
 &&\frac{\delta}{\delta
B^\alpha}S_{\mathrm{ext}}=D_\mu^{\alpha\beta}
(\mathcal{B})A_\mu^\beta+\xi B^\alpha, \\
\label{a25}
 &&D_\mu^{\alpha\beta}(\mathcal{B})\frac{\delta}{\delta
A_\mu^{\ast\beta}}
S_{\mathrm{ext}}-\frac{\delta}{\delta\overline{C}^\alpha}S_{\mathrm{ext}}=-
g\varepsilon^{\alpha\beta\gamma}A_\mu^\beta\theta_\mu^\gamma.
\eeq
The equation (\ref{a25}) means that the action $S_{ext}$ (\ref{a11})
depends on the variables $A_\mu^{\ast\alpha}$ and
$\overline{C}^\alpha$ only in combination
$A_\mu^{\ast\alpha}-D_\mu^{\alpha\beta}(\mathcal{B})
\overline{C}^\beta$ when $\theta_\mu^\beta=0$.

The action $S_{ext}$ can be represented in another useful form,
\beq
\label{a26} S_{ext}=\Gamma_{0|0}+\sum_{k=6}^{22}\Gamma_{0|k}+
\chi\sum_{k=1}^5\Gamma_{0|k}, \eeq where \beq \label{c3a}
&&\Gamma_{0|0}=\int dx\left(
B^{\alpha}D^{\alpha\beta}_{\mu}(\mathcal{B})
A^{\beta}_{\mu}+\frac{\xi}{2}B^{\alpha}B^{\alpha}+
g\theta_{\mu}^{\alpha}\varepsilon^{\alpha\beta\gamma}A_{\mu}^{\beta}
\overline{C}^{\gamma}\right), \\
\label{c17a}
&&\Gamma_{0|1}=\int dx\left[\mathcal{A}^{\ast\alpha}_\mu\mathcal{A}^\alpha_\mu
\right], \; \Gamma_{0|2}=\int dx\left[C^{\ast\alpha}C^\alpha\right],\;
\Gamma_{0|3}=\int dx\left[\psi^*_j\psi_j\right], \\
\label{c17b}
&&\Gamma_{0|4}=\int dx\left[\overline{\psi}^*_j\overline{\psi}_j\right], \;
\Gamma_{0|5}=\int dx\left[\varphi^{\ast\alpha}\varphi^\alpha\right], \;
\Gamma_{0|6}=\int dx\left[\mathcal{A}_\mu^{*\alpha}\theta_\mu^\alpha\right], \\
\label{d3a}
&&\Gamma_{0|7}=\int dx\left[\mathcal{A}_\mu^{*\gamma}
D_\mu^{\gamma\sigma}(\mathcal{B)}C^\sigma\right], \; \Gamma_{0|8}=\int dx
\left[g\mathcal{A}_\mu^{*\gamma}\varepsilon^{\gamma\alpha\sigma}
\mathcal{A}_\mu^\alpha C^\sigma\right], \\
\label{d8}
&&\Gamma_{0|9}=\int dx\left[\frac{g}{2}C^{*\alpha}
\varepsilon^{\alpha\beta\gamma}C^\beta C^\gamma\right], \;
\Gamma_{0|10}=-\int dx\left[g\psi^*t^\alpha\psi C^\alpha\right], \\
\label{d9}
&&\Gamma_{0|11}=\int dx\left[g\overline{\psi}t^\alpha\overline{\psi}^*
C^\alpha \right], \; \Gamma_{0|12}=-\int dx\left[g\varphi^{*\gamma}
\varepsilon^{\gamma\alpha\sigma}\varphi^\sigma C^\alpha\right], \\
\label{d14}
&&\Gamma_{0|13}=i\!\int\!\!dx\left[\overline{\psi}\gamma^\mu D_{\psi\mu}
(\mathcal{B})\psi\right], \; \Gamma_{0|14}=ig\!\int\!\!dx\left[\overline{\psi}
\gamma^\mu t^\alpha\mathcal{A}_\mu^\alpha\psi\right], \\
\label{d15a}
&&\Gamma_{0|15}=-m\!\int\!\!dx\left[\overline{\psi}\psi\right],\;
\Gamma_{0|16} =\int
dx\left[\frac{1}{2}(D_{\varphi\mu}^{\alpha\beta}(\mathcal{B})
\varphi^\beta)D_{\varphi\mu}^{\alpha\gamma}(\mathcal{B})\varphi^\gamma\right],
\\
\label{d15b} &&\Gamma_{0|17}=\!\!\int\!
dx\left[g\varepsilon^{\alpha\gamma
\sigma}(D_{\varphi\mu}^{\alpha\beta}(\mathcal{B})\varphi^\beta)
\mathcal{A}_\mu^\gamma\varphi^\sigma\right],\;\Gamma_{0|18}\!=\!\int\!
dx
\left[\frac{g^2}{2}E_{\alpha\beta}^\rho\mathcal{A}_\mu^\alpha\varphi^\beta
E_{\gamma\sigma}^\rho\mathcal{A}_\mu^\gamma\varphi^\sigma\right]\!, \\
\label{d15c}
&&\Gamma_{0|19}=-\frac{M^2}{2}\!\!\int\!\!dx\varphi^2, \; \Gamma_{0|20}=-
\frac{\lambda}{4}\!\int\!dx\varphi^4,\;\Gamma_{0|21}=i\vartheta\int dx\left[
\varphi^\alpha\overline{\psi}t^\alpha\psi\right], \\
\label{d16a}
&&\Gamma_{0|22}=-\frac{1}{4}\int dx\left[G_{\mu\nu}^\alpha
(\mathcal{A}+\mathcal{B})G_{\mu\nu}^\alpha(\mathcal{A}+\mathcal{B})\right],
\eeq
and the notations
\beq
\label{a39}
\mathcal{A}^{*\alpha}_\mu=A^{*\alpha}_{\mu}-D^{\alpha\beta}_{\mu}({\mathcal B})
\overline{C}^{\beta},\quad \mathcal{A}^{\alpha}_{\mu}=A^{\alpha}_{\mu}
\eeq
are used. Note that the functionals  $\Gamma_{0|0} - \Gamma_{0|21}$ are
homogeneous with respect to  fields $\Omega$ and antifields $\Omega^*$ while
the functional  $\Gamma_{0|22}$ does not obey this property and in symbolic
notation has the form
\beq
\label{17}
\Gamma_{0|22}=\sum_{l=0}^4\Gamma_{0|22,l}, \; \Gamma_{0|22,l}=
\int dx\Big[T_l(\mathcal{B})\mathcal{A}^l\Big],
\eeq
where $\Gamma_{0|22,l}$ is  homogeneous functional with respect to  variable
$\mathcal{A}$ of order $l$ and $T_l(\mathcal{B})$ is a tensor (differential
operator) with $l$ gauge and $l$ Lorentz indices.  In what follow we do not
need in explicit representation of quantities $T_l(\mathcal{B})$.

Later on, we will see that the generating functional of vertex
functions (effective action),  counterterms  and renormalized action
satisfy the same equations (\ref{a22}), (\ref{a23}), (\ref{a23a}),
(\ref{a24}), (\ref{a25}), as the action $S_{\mathrm{ext}}$ does.
Moreover, the counterterms and the renormalized action are linear
combinations of the same vertices as the action $S_{\mathrm{ext}}$
is. Explicit form of these vertices will be given in the next
section.

Now we give the table of "quantum" numbers of fields, antifields, auxiliary
fields and constant parameters used in construction of $S_{ext}$:
\begin{center}
\begin{tabular}[c]{|c|c|c|c|c|c|c|c|c|c|c|c|c|c|c|}
\hline
Quantity & $A,{\cal B}$ & $\psi,\overline{\psi}$ & $\varphi$, &
$C,\overline{C}$ & $B$ & $\xi$ & $\theta $ & $A^*$ &
$\psi^*,\overline{\psi}^*$ & $\varphi^*$ & $C^*$ & $dx(\equiv d^4x)$ &
$\pa_{x}$ & $\chi$  \\
\hline
$\varepsilon$ & 0& 1& 0& 1& 0& 0& 1& 1& 0& 1& 0& 0& 0& 1  \\
\hline
{\rm gh} & 0& 0& 0& 1,-1& 0& 0& 1& -1& -1& -1& -2& 0& 0& 1  \\
\hline
dim & 1& 3/2& 1& 1& 2& 0& 2& 2& 3/2& 2& 2& -4& 1& 1  \\
\hline
$\varepsilon_f$ & 0& 1,-1& 0& 0& 0& 0& 0& 0& -1,1& 0& 0& 0& 0& 0 \\
\hline
\end{tabular}
\end{center}
where $"\varepsilon"$ describes the Grassmann parity, the symbol
 $"{\rm gh}"$ is used to denote the ghost number,
 $"{\rm dim}"$ means the canonical dimension and   $"\varepsilon_f"$ is the
fermionic number. Using the table of "quantum" numbers it is easy to establish
the quantum numbers of any quantities met in the text.

\section{General structure of renormalized action}
\noindent

It will be  shown below that the renormalized action is a local
functional of field variables with quantum numbers of the action
${\cal S}_{ext}$ (\ref{a11}), which satisfies the same equations
(\ref{a22}) - (\ref{a25}) as the action $S_{ext}$ does. In this
section, we  find a general solution to the equations (\ref{a22}) -
(\ref{a25}) under the conditions pointed out above.

So, let $P$ be a functional of the form \beq \label{c1} P=\int dx
P(x), \eeq where $P(x)$ is a local polynomial of all variables
$Q,Q^*,\overline{C}, B, {\cal B},\xi,\theta,\chi$ with ${\rm
dim}(P(x))=4$.  Let the functional $P$ satisfies the equations
(\ref{a22}) - (\ref{a25}) (with the substitution $S_{ext} \to P$), and
we  represent it  in the form \beq \label{c2}
P=\Gamma_{0|0}+P^{(1)}+\chi P^{(2)}, \eeq where the functional
$\Gamma_{0|0}$ is defined in (\ref{c3a}), the  functionals $P^{(1)},
P^{(2)}$ do not depend on  $\chi$ and obey the properties
\beq
\label{c4a} &&\varepsilon (P^{(1)})=0,\quad {\rm gh}(P^{(1)})=0,
\quad
{\rm dim}(P^{(1)})=0,\quad \varepsilon_f((P^{(1)}))=0, \\
\label{c4b}
&&\varepsilon(P^{(2)})=1,\quad {\rm gh}(P^{(2)})=-1,\quad
{\rm dim}(P^{(2)})=-1,\quad \varepsilon_f((P^{(2)}))=0.
\eeq
From the equation (\ref{a24}) for $P$ and the presentation (\ref{c2}) it
follows that $P^{(1)}$ and $P^{(2)}$ do not depend on fields $B^{\alpha}$,
\beq
\label{c5}
 P^{(k)}=P^{(k)}(Q,Q^{\ast },\overline{C},\mathcal{B},\xi ,\theta ),\ k=1,2,
\eeq
With the help of new variables $\mathcal{A}_\mu^{\ast\alpha}(x),\;
{\mathcal A^{\alpha}}_{\mu}$ (\ref{a39}), we define new functionals
$\tilde{P}^{(k)}$ by the rule
\beq
\label{c7}
\tilde{P}^{(k)}=\tilde{P}^{(k)}(\Omega,\Omega^{*},\mathcal{B},\overline{C},
\xi,\theta)=P^{(k)}(Q,Q^*,\mathcal{B},\overline{C},\xi,\theta)|_{A^*\to
\mathcal{A}^*+D(\mathcal{B})\overline{C}},\quad k=1,2,
\eeq
and find that $\tilde{P}^{(k)}$ do not depend on fields $\overline{C}^\alpha$,
\beq
\label{c8}
\tilde{P}^{(k)}=\tilde{P}^{(k)}(\Omega,\Omega^\ast,\mathcal{B},\xi,\theta).
\eeq
 Omitting the indices of all variables in relations (\ref{c7}) and (\ref{c8}), the
 following notations
\beq
\label{c9}
\Omega
=\{\mathcal{A},\psi,\overline{\psi},\varphi,C\},\quad
\Omega^\ast=\{\mathcal{A}^*,\psi^*,\overline{\psi}^*,\varphi^*,C^*\}
\eeq
 are used. Independence of the functionals $\tilde{P}^{(k)}$ of
the fields $\overline{C}^\alpha$ and the relations
\beq
\nonumber
&&P^{(k)}\int
dx\Big[\frac{\overleftarrow{\delta}}{\delta\mathcal{B}_\mu^\beta}
D_\mu^{\beta\alpha}(\mathcal{B})+g\varepsilon^{\beta\gamma\alpha}\Big
(\frac{\overleftarrow{\delta}}{\delta\overline{C}^\beta}\overline{C}^\gamma+
\frac{\overleftarrow{\delta}}{\delta A_\mu^{*\beta}}
A_\mu^{\ast\gamma}\Big)\Big]= \\
\label{c10} &&\qquad =\tilde{P}^{(k)}\int
dx\Big[\frac{\overleftarrow{\delta}}
{\delta\mathcal{B}_\mu^\beta}D_\mu^{\beta\alpha}(\mathcal{B})+
g\varepsilon^{\beta\gamma\alpha}\frac{\overleftarrow{\delta}}
{\delta\mathcal{A}_\mu^{\ast\beta}}\mathcal{A}_\mu^{\ast\gamma}\Big],
\eeq
allow  us to write  down the following set of equations for
$\tilde{P}^{(k)}$
\beq \label{c11} &&\qquad \qquad \int
dx\Big[\tilde{P}^{(1)}\frac{\overleftarrow{\delta}}{\delta
\Omega}\frac{\delta}{\delta\Omega^*}\tilde{P}^{(1)}-\theta_\mu^\alpha
\frac{\delta}{\delta\mathcal{B}_\mu^\alpha}\tilde{P}^{(1)}\Big] =0, \\
\nonumber
&&2\xi \frac{\partial}{\partial\xi}\tilde{P}^{(1)}=\int dx\Big[
\tilde{P}^{(1)}\Big(\frac{\overleftarrow{\delta}}{\delta\Omega}
\frac{\delta}{\delta\Omega^*}-\frac{\overleftarrow{\delta}}
{\delta\Omega^*}\frac{\delta}{\delta\Omega}\Big)\tilde{P}^{(2)}-
\theta_\mu^\alpha\frac{\delta}{\delta\mathcal{B}_\mu^\alpha}
\tilde{P}^{(2)}\Big] + \\
\label{c12}
&&\qquad \qquad\qquad +\int dx\Big[\Big(\Omega^*\frac{\delta}{\delta\Omega^*}
-\Omega\frac{\delta}{\delta\Omega}\Big)\tilde{P}^{(1)}\Big ],
\eeq
\beq
\label{c13}
\tilde{P}^{(k)}\overleftarrow{\tilde{h}^{\alpha}}\omega_{\alpha}=0,\ k=1,2,
\eeq
where
\beq
&&\overleftarrow{\tilde{h}^\alpha}\omega_\alpha=\int dx\Big\{\Big[
\frac{\overleftarrow{\delta}}{\delta\mathcal{B}_\mu^\beta}%
D_\mu^{\beta\alpha}(\mathcal{B})+g\varepsilon^{\beta\gamma\alpha}\Big(
\frac{\overleftarrow{\delta}}{\delta\mathcal{A}_\mu^\beta}\mathcal{A}%
_{\mu}^\gamma+\frac{\overleftarrow{\delta}}{\delta\varphi^\beta}
\varphi^\gamma+\frac{\overleftarrow{\delta}}{\delta C^\beta}C^\gamma
\Big) +  \notag \\
&&\qquad\qquad+g\varepsilon^{\beta\gamma\alpha}\Big(\frac{\overleftarrow
{\delta}}{\delta\mathcal{A}_\mu^{\ast\beta}}\mathcal{A}_\mu^{\ast\gamma}+
\frac{\overleftarrow{\delta}}{\delta \varphi^{\ast\beta}}
\varphi^{\ast\gamma}+\frac{\overleftarrow{\delta}}{\delta C^{\ast\beta}}
C^{\ast\gamma }+\frac{\overleftarrow{\delta}}{\delta\theta_\mu^\beta}
\theta_\mu^\gamma\Big)-  \notag \\
\label{c14}
&&\qquad\qquad-gt_{jk}^\alpha\Big(\frac{\overleftarrow{\delta}}{\delta\psi_j}
\psi_k+\frac{\overleftarrow{\delta}}{\delta\overline{\psi}_j^*}
\overline{\psi}_k^*\Big)+g\Big(\frac{\overleftarrow{\delta}}
{\delta\overline{\psi}_j}\overline{\psi}_k+\frac{\overleftarrow{\delta}}
{\delta\psi_j^*}\psi_k^*\Big)t_{kj}^\alpha\Big]\omega_\alpha\Big\}.
\eeq

When studying the structure of functionals $\tilde{P}^{(k)}$ and in
further research, it is helpful to have a consequence from equation
(\ref{c13}) that corresponds to the case when $\omega_{\alpha}={\rm
const}$,
\beq
\label{c15} \tilde{P}^{(k)}\overleftarrow{T^{\alpha
}}=0,\ k=1,2, \eeq where \beq &&\overleftarrow{T^{\alpha }}=\int
dx\Big\{\varepsilon^ {\beta\gamma\alpha}\Big(
\frac{\overleftarrow{\delta}}
{\delta\mathcal{B}_\mu^\beta}\mathcal{B}_\mu^\gamma+
\frac{\overleftarrow{\delta}}{\delta\mathcal{A}_\mu^\beta}
\mathcal{A}_\mu^\gamma+\frac{\overleftarrow{\delta}}{\delta\varphi^\beta}
\varphi^\gamma+\frac{\overleftarrow{\delta}}{\delta
C^\beta}C^\gamma\Big)
+ \notag \\
&&\qquad\qquad+\varepsilon^{\beta\gamma\alpha}\Big(\frac{\overleftarrow{\delta}}
{\delta\mathcal{A}_\mu^{\ast\beta}}\mathcal{A}_\mu^{\ast\gamma}+
\frac{\overleftarrow{\delta}}{\delta \varphi^{\ast\beta}}\varphi^{\ast\gamma}+
\frac{\overleftarrow{\delta}}{\delta C^{\ast\beta}}C^{\ast\gamma}+
\frac{\overleftarrow{\delta}}{\delta\theta_\mu^\beta}
\theta_\mu^\gamma\Big)-  \notag \\
\label{c16}
&&\qquad\qquad-t_{jk}^\alpha\Big(\frac{\overleftarrow{\delta}}{\delta\psi_j}
\psi_k+\frac{\overleftarrow{\delta}}{\delta\overline{\psi}_j^*}
\overline{\psi}_k^*\Big)+\Big(\frac{\overleftarrow{\delta}}{\delta
\overline{\psi}_j}\overline{\psi}_k+\frac{\overleftarrow{\delta}}
{\delta\psi_j^*}\psi_k^*\Big)t_{kj}^\alpha\Big\}.
\eeq
We refer to
the equation of the form (\ref{c15}) as the ones of the $T$-symmetry
for the corresponding functionals.

Using the properties of the functional $\tilde{P}^{(2)}$: its locality and
(\ref{c4b}) as well as axial symmetry, Poincare- and $T$-symmetries, we find
the general representation  of $\tilde{P}^{(2)}$
\beq
\label{c17}
\tilde{P}^{(2)}=\sum_{k=1}^5Z_k\Gamma_{0|k}+\int dx\left[Z_1^\prime
\mathcal{A}^{\ast\alpha}_\mu\mathcal{B}^\alpha_\mu\right],
\eeq
where $Z_i, i=1,2,3,4,5$ and $Z_{1}^{\prime }$ are arbitrary constants and the
functionals $\Gamma_{0|k}$, $k=1,2,3,4,5$ were introduced in (\ref{c17a}),
(\ref{c17b}). Later on any  quantities $"Z"$ with any set of indices do not
depend on coordinates $x$ and field variables.

Further, when using  the equation (\ref{c13}) for $\tilde{P}^{(2)}$, we get
that $Z_1^\prime=0$. The final expression for the $\tilde{P}^{(2)}$ has the
form
\beq
\label{c18}
\tilde{P}^{(2)}=\sum_{k=1}^5Z_k\Gamma_{0|k},
\eeq

Notice that the functional $\tilde{P}^{(2)}$ does not depend on the fields
$\theta_\mu^\alpha$ and $\mathcal{B}_\mu^\alpha$. By taking  (\ref{c18}) into
account the equation (\ref{c12}) reduces to the following one
\beq
\label{c19}
&&\hspace{4,5cm}2\xi\frac{\partial}{\partial\xi}\tilde{P}^{(1)}=
\hat{L}(Z)\tilde{P}^{(1)}, \\
\nonumber
&&\hat{L}(Z)=\int dx\Big[( Z_1-1)\Big(
\mathcal{A}^\alpha_\mu\frac{\delta}{\delta\mathcal{A}^\alpha_\mu}-
\mathcal{A}^{*\alpha}_\mu\frac{\delta}{\delta\mathcal{A}_\mu^{*\alpha}}\Big)+
(Z_2-1)\Big(C^{\alpha}\frac{\delta}{\delta C^{\alpha}}-
C^{*\alpha}\frac{\delta}{\delta C^{*\alpha}}\Big)+ \\
\nonumber
&&\qquad\qquad+(Z_3-1)\Big(\psi_j\frac{\delta}{\delta\psi_j}-\psi^*_j
\frac{\delta}{\delta{\psi}^*_j}\Big)+(Z_4-1)\Big(\overline{\psi}_j\frac{\delta}
{\delta\overline{\psi}_j}-\overline{\psi}^*_j\frac{\delta}
{\delta\overline{\psi}^*_j}\Big)+ \\
\label{c19a}
&&\qquad\qquad+(Z_5-1)\Big(\varphi^{\alpha}\frac{\delta}{\delta\varphi^\alpha}-
\varphi^{*\alpha}\frac{\delta}{\delta\varphi^{*\alpha}}\Big)\Big],
\eeq
describing the dependence of renormalization constants on the gauge parameter
$\xi$. We refer to the equation ({\ref{c11}}) as the extended master-equation
and to ({\ref{c19}}) as the gauge dependence equation.

\subsection{Solution to the extended master-equation}
\noindent
Let us consider a solution to the extended master-equation
({\ref{c11}}) for the functional $\tilde{P}^{(1)}$ presented in the
form
\beq
\label{d1}
\tilde{P}^{(1)}=\tilde{P}_\theta^{(1)}+\tilde{P}_{\Omega^*}^{(1)}+
\tilde{P}_\psi^{(1)}+\tilde{P}_\varphi^{(1)}+\tilde{P}_{\psi\varphi}^{(1)}
+\tilde{P}_\mathcal{AB}^{(1)}.
\eeq
Functional
$\tilde{P}_\theta^{(1)}$ is written as
\beq
\label{d2}
\tilde{P}_\theta^{(1)}=\int dx\theta_\mu^\alpha(x)
\tilde{P}_{\theta\mu}^\alpha(x),
\eeq
and functionals
$\tilde{P}_{\Omega^*}^{(1)}$, $\tilde{P}_\psi^{(1)}$,
$\tilde{P}_\varphi^{(1)}$, $\tilde{P}_{\psi\varphi}^{(1)}$,
$\tilde{P}_\mathcal{AB}^{(1)}$ do not depend on the fields
$\theta^{\alpha}_{\mu}$. Taking into account the properties ${\rm
dim}(\tilde{P}_{\theta\mu}^\alpha)=2$, ${\rm
gh}(\tilde{P}^{\alpha}_{\mu\theta })=-1$, $\varepsilon
(\tilde{P}_{\theta\mu}^\alpha)=1$,
$\varepsilon_{f}(\tilde{P}_{\theta\mu}^\alpha)=0$ as well as the
Poincare- and $T$-symmetries  of $\tilde{P}_\theta^{(1)}$, we find
that
\beq
\label{d3}
\tilde{P}_{\theta\mu}^\alpha(x)=-Z_6\mathcal{A}_\mu^{*\alpha}(x),
\quad \tilde{P}_\theta^{(1)}=Z_6\Gamma_{0|6},
\eeq
where  $Z_6$ is an arbitrary constant.

Functional $\tilde{P}_{\Omega^*}^{(1)}$ (\ref{c9}) is linear in
antifields $\Omega^*$ and functionals $\tilde{P}_\psi^{(1)}$,
$\tilde{P}_\varphi^{(1)}$, $\tilde{P}_{\psi\varphi}^{(1)}$,
$\tilde{P}_\mathcal{AB}^{(1)}$ do not depend on antifields
$\Omega^*$. The functional  $\tilde{P}_{\Omega^*}^{(1)}$  can be
presented in the form
\beq
\label{d4}
\tilde{P}_{\Omega^*}^{(1)}=\tilde{P}_{\mathcal{A}^*}^{(1)}+
\tilde{P}_{C^*}^{(1)}+\tilde{P}_{\psi^*}^{(1)}+
\tilde{P}_{\overline{\psi}^*}^{(1)}+\tilde{P}_{\varphi^*}^{(1)}.
\eeq
Using the arguments similar to those that led us to establish
the form of the functional $\tilde{P}_\theta^{(1)}$ (\ref{d3}), we
obtain
\beq
\label{d5}
&&\tilde{P}_{\mathcal{A}^*}^{(1)}=Z_7\Gamma_{0|7}+Z_8\Gamma_{0|8},
\quad
\tilde{P}_{C^*}^{(1)}=Z_9\Gamma_{0|9}, \\
\label{d6}
&&\tilde{P}_{\psi^*}^{(1)}=Z_{10}\Gamma_{0|10}, \quad
\tilde{P}_{\overline{\psi}^*}^{(1)}=Z_{11}\Gamma_{0|11}, \quad
\tilde{P}_{\varphi^*}^{(1)}=Z_{12}\Gamma_{0|12}.
\eeq

Functional $\tilde{P}_\psi^{(1)}$ depends on the variables $\Psi$,
$\mathcal{A}$, $\mathcal{B}$, and is quadratic in $\Psi$. Taking
into account the axial symmetry as well as the Poincare- and
$\mathcal{B}$-gauge invariance, we find the general structure of
$\tilde{P}_\psi^{(1)}$,
\beq
\label{d13}
\tilde{P}_\psi^{(1)}=Z_{13}\Gamma_{0|13}+Z_{14}\Gamma_{0|14}+
Z_{15}\Gamma_{0|15}. \eeq

Functional $\tilde{P}_\varphi^{(1)}$ depends on the variables
$\varphi$, $\mathcal{A}$, $\mathcal{B}$, and vanishes when
$\varphi=0$. Taking into account the Poincare- and the
$\mathcal{B}$-gauge invariance, we establish the general form of the
functional $\tilde{P}_\varphi^{(1)}$,
\beq
\label{d15}
\tilde{P}_\varphi^{(1)}=Z_{16}\Gamma_{0|16}+Z_{17}\Gamma_{0|17}+
\tilde{P}_{\varphi18}^{(1)}+Z_{19}\Gamma_{0|19}+Z_{20}\Gamma_{0|20},
\eeq
where
\beq
\label{d15a2}
\tilde{P}_{\varphi18}^{(1)}=\!\int
dx\!\left[\frac{g^2}{2}Z_{18}^
{\alpha\beta\gamma\sigma}\mathcal{A}_\mu^\alpha\varphi^\beta
\mathcal{A}_\mu^\gamma\varphi^\sigma\right], \quad
Z_{18}^{\alpha\beta\gamma\sigma}=Z_{18}^{\gamma\beta\alpha\sigma}=
Z_{18}^{\alpha\sigma\gamma\beta}.
 \eeq
 If the condition
\beq
\label{d15d}
\int dx\left[Z_{18}^{\alpha\beta\gamma\sigma}\mathcal{A}_\mu^\alpha
\varphi^\beta\mathcal{A}_\mu^\gamma\varphi^\sigma\right]=Z_{18}\int
dx\left[\sum_{\sigma=1}^3E_{\sigma\beta}^\alpha E_{\sigma\rho}^\gamma
\mathcal{A}_\mu^\alpha\varphi^\beta\mathcal{A}_\mu^\gamma
\varphi^\rho\right]
\eeq
fulfils then the functional $\tilde{P}_{\varphi18}^{(1)}$ satisfies the
equation (\ref{c13}).

Functional $\tilde{P}_{\psi\varphi}^{(1)}$ is an interaction vertex
of the fields $\varphi$  and $\Psi$.  Taking into account the axial,
Poincare- and $T$-symmetries, this functional has the following
general form:
\beq
\label{d16}
\tilde{P}_{\psi\varphi}^{(1)}=Z_{21}\Gamma_{0|21}. \eeq The
functional $\tilde{P}_{\psi\varphi}^{(1)}$ (\ref{d16}) satisfies the
equation (\ref{c13}).

Substitute into the equation (\ref{c11}) the representation for the
functional $\tilde{P}^{(1)}$ in the form  (\ref{d1}) where the terms
$\tilde{P}_{\theta}^{(1)}$, $\tilde{P}_{\Omega^*}^{(1)}$,
$\tilde{P}_\psi^{(1)}$, $\tilde{P}_\varphi^{(1)}$ and
$\tilde{P}_{\psi\varphi}^{(1)}$ are described by the relations
(\ref{d3}), (\ref{d4}) - (\ref{d6}), (\ref{d13}), (\ref{d15}) -
(\ref{d15a}) and (\ref{d16}), respectively, and take into account
that $\tilde{P}_{\mathcal{AB}}^{(1)}=
\tilde{P}_{\mathcal{AB}}^{(1)}(\mathcal{A,B})$.  Then, a solution to
the equation (\ref{c11}) is reduced to solutions to the
sub-equations which follow from the requirement for independent
polynomial structures appearing in the left-hand side of the
equation (\ref{c11}), to be equal to zero. In their turn, these
sub-equations are reduced to algebraic equations for coefficients
$"Z"$ or, in two cases, to variational differential equations for
the functional $\tilde{P}_{\mathcal{AB}}^{(1)}(\mathcal{A,B})$.

We explain the results obtained by using an example for the block
\beq
\label{d17a1}
&&\theta\mathcal{A}^*C\;\;\Rightarrow \int
dx\Big[Z_6Z_8\Gamma_{0|8}
\frac{\overleftarrow{\delta}}{\delta\mathcal{A}_\mu^\alpha}
\frac{\delta}{\delta\mathcal{A^{*\alpha}_\mu}}\Gamma_{0|6}-
Z_7\theta_\mu^\alpha\frac{\delta}{\delta\mathcal{B}_\mu^\alpha}\Gamma_{0|7}
\Big] =0 \; \Rightarrow \\
\label{d17a2} &&\qquad\qquad\quad  Z_8=\frac{Z_7}{Z_6}.
\eeq
This block should be understood as the following: the requirement for the
structure $\theta\mathcal{A}^*C$ to be equal to zero leads to
equation (\ref{d17a1}), from which it follows the relation
(\ref{d17a2}).

Further
\beq
\nonumber
&&\theta\overline{\psi}\gamma^\mu\psi\;\;\Rightarrow \int dx\Big[Z_6Z_{14}
\Gamma_{0|14}\frac{\overleftarrow{\delta}}{\delta\mathcal{A}_\mu^\alpha}
\frac{\delta}{\delta\mathcal{A^{*\alpha}_\mu}}\Gamma_{0|6}-Z_{13}
\theta_\mu^\alpha\frac{\delta}{\delta\mathcal{B}_\mu^\alpha}\Gamma_{0|13}
\Big]=0 \; \Rightarrow \\
\label{d17b}
&&\qquad\qquad\quad Z_{14}=\frac{Z_{13}}{Z_6};\\
\nonumber
&&\theta\varphi(D(\mathcal{B})\varphi)\;\; \Rightarrow \int dx\Big[Z_6Z_{17}
\Gamma_{0|17}\frac{\overleftarrow{\delta}}{\delta\mathcal{A}_\mu^\alpha}
\frac{\delta}{\delta\mathcal{A^{*\alpha}_\mu}}\Gamma_{0|6}-
Z_{16}\theta_\mu^\alpha\frac{\delta}{\delta\mathcal{B}_\mu^\alpha}\Gamma_{0|16}
\Big]=0 \; \Rightarrow \\
\label{d17c}
&&\qquad\qquad\qquad  Z_{17}=\frac{Z_{16}}{Z_6};\\
\nonumber
&&\theta\mathcal{A}\varphi\varphi\;\;\Rightarrow \int dx\Big[Z_6\tilde{P}_
{\varphi18}^{(1)}\frac{\overleftarrow{\delta}}{\delta\mathcal{A}_\mu^\alpha}
\frac{\delta}{\delta\mathcal{A^{*\alpha}_\mu}}\Gamma_{0|6}-Z_{17}
\theta_\mu^\alpha\frac{\delta}{\delta\mathcal{B}_\mu^\alpha}\Gamma_{0|17}
\Big]=0 \; \Rightarrow \\
\label{d17d}
&&\qquad\qquad\quad Z_{18}^{\alpha\beta\gamma\sigma}\varphi^\beta
\mathcal{A}_\mu^\gamma\varphi^\sigma=\frac{Z_{17}}{Z_6}E_{\alpha\beta}^\rho
E_{\gamma\sigma}^\rho\varphi^\beta\mathcal{A}_\mu^\gamma\varphi^\sigma.
\eeq

Multiplying the equality (\ref{d17d})  by $\mathcal{A}_\mu^\alpha$
and integrating then over $x$, we find that the relation
(\ref{d15d}) is satisfied and the functional
$\tilde{P}_{\varphi18}^{(1)}$ reads
\beq
 \label{d17e}
\tilde{P}_{\varphi18}^{(1)}=Z_{18}\Gamma_{0|18}, \quad
Z_{18}=\frac{Z_{17}}{Z_6}=\frac{Z_{16}}{Z_6^2}. \eeq \beq \nonumber
&&\theta\mathcal{A}^n\mathcal{B}^k\;\;\Rightarrow \int dx\Big[Z_6
\tilde{P}_{\mathcal{AB}}^{(1)}\frac{\overleftarrow{\delta}}
{\delta\mathcal{A}_\mu^\alpha}\frac{\delta}{\delta\mathcal{A^{*\alpha}_\mu}}
\Gamma_{0|6}-\theta_\mu^\alpha\frac{\delta}{\delta\mathcal{B}_\mu^\alpha}
\tilde{P}_{\mathcal{AB}}^{(1)}\Big]=0 \; \Rightarrow \\
\label{d17f}
&&\qquad\Big(Z_6\frac{\delta}{\delta\mathcal{A}_\mu^\alpha}-\frac{\delta}
{\delta\mathcal{B}_\mu^\alpha}\Big)\tilde{P}_{\mathcal{AB}}^{(1)}=0\
\Rightarrow \tilde{P}_{\mathcal{AB}}^{(1)}=\tilde{P}_{\mathcal{AB}}^{(1)}(U),\
U\!=U(Z_6)=\mathcal{B}\!+\!\frac{1}{Z_6}\mathcal{A};\\
\nonumber
&&\mathcal{A}^*\mathcal{A}CC\;\;\Rightarrow \int dx\Big[Z_8^2\Gamma_{0|8}
\frac{\overleftarrow{\delta}}{\delta\mathcal{A}_\mu^\alpha}\frac{\delta}
{\delta\mathcal{A^{*\alpha}_\mu}}\Gamma_{0|8}+Z_8 Z_9\Gamma_{0|8}
\frac{\overleftarrow{\delta}}{\delta C^\alpha}\frac{\delta}{\delta C^{*\alpha}}
\Gamma_{0|9}\Big]=0 \; \Rightarrow \\
\label{d18a}
&&\qquad\qquad\quad Z_9=Z_8=\frac{Z_7}{Z_6};\\
\nonumber
&&\psi^*\psi CC \; \Rightarrow \int dx\Big[Z_{10}^2\Gamma_{0|10}\frac
{\overleftarrow{\delta}}{\delta\psi_j}\frac{\delta}{\delta\psi_j^*}
\Gamma_{0|10}+Z_{10}Z_9\Gamma_{0|10}\frac{\overleftarrow{\delta}}
{\delta C^\alpha}\frac{\delta}{\delta C^{*\alpha}}\Gamma_{0|9}\Big]=0
\; \Rightarrow \\
\label{d18ab}
&&\qquad\qquad\quad Z_{10}=Z_9=\frac{Z_7}{Z_6};\\
\nonumber
&&\overline{\psi}\overline{\psi}^*CC \; \Rightarrow \int dx\Big[Z_{11}^2
\Gamma_{0|11}\frac{\overleftarrow{\delta}}{\delta\overline{\psi}_j}
\frac{\delta}{\delta\overline{\psi}_j^*}\Gamma_{0|11}+Z_{11}Z_9\Gamma_{0|11}
\frac{\overleftarrow{\delta}}{\delta C^\alpha}\frac{\delta}{\delta C^{*\alpha}}
\Gamma_{0|9}\Big]=0 \; \Rightarrow \\
\label{d18ac}
&&\qquad\qquad\quad Z_{11}=Z_9=\frac{Z_7}{Z_6};
\eeq
\beq
\nonumber
&&\overline{\psi}\psi\varphi C\;\;\Rightarrow \int dx\Big[Z_{10}Z_{21}
\Gamma_{0|21}\frac{\overleftarrow{\delta}}{\delta\psi_j}\frac{\delta}
{\delta\psi_j^*}\Gamma_{0|10}+Z_{11}Z_{21}\Gamma_{0|21}\frac
{\overleftarrow{\delta}}{\delta\overline{\psi}_j}\frac{\delta}
{\delta\overline{\psi}_j^*}\Gamma_{0|11}+ \\
\nonumber
&&\qquad\qquad\quad +Z_{12}Z_{21}\Gamma_{0|21}\frac{\overleftarrow{\delta}}
{\delta\varphi^\alpha}\frac{\delta}{\delta\varphi^{*\alpha}}\Gamma_{0|12}
\Big]=0 \; \Rightarrow \\
\label{d18c}
&&\qquad\qquad\quad Z_{12}=Z_{10}=\frac{Z_7}{Z_6};\\
\nonumber
&&\mathcal{A}^n\mathcal{B}^kC \;\Rightarrow \int dx\Big[\tilde{P}_
{\mathcal{AB}}^{(1)}\frac{\overleftarrow{\delta}}{\delta\mathcal{A}_\mu^\alpha}
\frac{\delta}{\delta\mathcal{A^{*\alpha}_\mu}}\big(Z_7\Gamma_{0|7}+
Z_8\Gamma_{0|8}\big)\Big]=0 \; \Rightarrow \\
&&\qquad\qquad D_\mu^{\alpha\beta}(U)\frac{\delta}{\delta U_\mu^\beta}
\tilde{P}_{\mathcal{AB}}^{(1)}(U)=0\;\Rightarrow
\tilde{P}_{\mathcal{AB}}^{(1)}=Z_{22}\Gamma_{0|22}(Z_6), \\
\label{d18d}
&&\qquad\qquad\Gamma_{0|22}(Z_6)=-\frac{1}{4}\int dx\left[
G_{\mu\nu}^\alpha(U(Z_6))G_{\mu\nu}^\alpha(U(Z_6))\right].
\eeq

Requirement for the rest structures in the left-hand side of the equation
(\ref{c11})  to be equal to zero is satisfied identically.

Notice that the functional $\tilde{P}^{(1)}$ can be represented as a
linear combination of independent polynomials $\Gamma_{0|k}$,
\beq
\label{d20}
\tilde{P}^{(1)}=\sum_{k=6}^{22}Z_k\Gamma_{0|k}.
\eeq

\subsection{Solution to the gauge dependence equation}

Consider now a solution to the equation  (\ref{c19}) describing the
dependence of constants $"Z"$ on the gauge parameter $\xi$ appearing
in the general solution to the extended master-equation (\ref{c11})
for the functional $\tilde{P}^{(1)}$. For the functional
$\tilde{P}^{(1)}$ we will use the representation (\ref{d20}). Notice
that every functional $\Gamma_{0|k}$, $k=6,...,21$ is eigen for the
operator $\hat{L}$ (\ref{c19a}),
\beq
\label{e0}
\hat{L}\Gamma_{0|k}=\zeta_k\Gamma_{0|k},
\eeq
where $\zeta_k$ are
eigenvalues of the operator $\hat{L}$. It means that the equation
(\ref{c19}) reduces to the set of equations of the form
\beq
\label{e01} 2\xi\dot{Z}_k=\zeta_k, \; k=6,...,21.
\eeq
In relations
(\ref{e01}) and later on for any quantity $I=I(\xi ,...)$ we use the
notation
\beq
\label{e02}
\dot{I}\equiv\frac{\partial}{\partial\xi}I.
\eeq

The equations (\ref{e01}) for independent constants $Z_k$ lead to the
following relations and consequences,
\beq
\label{e6}
&&k=6, \quad 2\xi\dot{Z}_6=Z_6(1-Z_1) \; \Rightarrow \; Z_1=1-2\xi
\frac{\dot{Z}_6}{Z_6},\\
\label{e7}
&&k=7,\quad
2\xi\dot{Z}_7=Z_7(Z_2-Z_1) \; \Rightarrow \; Z_2=1+
2\xi\Big(\frac{\dot{Z}_7}{Z_7}-\frac{\dot{Z}_6}{Z_6}\Big),\\
\label{e13} &&k=13,\quad 2\xi\dot{Z}_{13}=Z_{13}(Z_3+Z_4-2) \;
\Rightarrow \; Z_4=2+
2\xi\frac{\dot{Z}_{13}}{Z_{13}}-Z_3,\\
\label{e15}
&&k=15,\quad\!\!
2\xi\dot{Z}_{15}=Z_{15}(Z_3+Z_4-2) \; \Rightarrow \; \dot{Z}_{23}=0, \;
Z_{15}=Z_{13}Z_{23},\\
\label{e16}
&&k=16,\quad \xi\dot{Z}_{16}=Z_{16}(Z_5-1) \; \Rightarrow \;
Z_5=1+\xi\frac{\dot{Z}_{16}}{Z_{16}},\\
\label{e19}
&&k=19,\quad
\xi\dot{Z}_{19}=Z_{19}(Z_5-1) \; \Rightarrow \; \dot{Z}_{24}=0, \;
Z_{19}=Z_{16}Z_{24},\\
\label{e20}
&&k=20,\quad
\xi\dot{Z}_{20}=2Z_{20}(Z_5-1) \; \Rightarrow \; \dot{Z}_{25}=0, \;
Z_{20}=Z_{16}^2Z_{25},\\
\label{e21}
&&k=21,\quad
\label{e21a}
2\xi\dot{Z}_{21}=Z_{21}(Z_3+Z_4+Z_5-3) \; \Rightarrow \; \dot{Z}_{26}=0, \;
Z_{21}=Z_{13}Z_{16}^{1/2}Z_{26}.
\eeq

As to the functional $\Gamma_{0|22}(Z_6)$, let us represent it in
the form
\beq
\label{e22l}
\Gamma_{0|22}(Z_6)=\sum_{l=0}^{4}Z_6^{-l}\Gamma_{0|22,l}.
\eeq
Then,
from the equation (\ref{c19}), it follows
\beq
\label{e22}
&& l=0,\quad \dot{Z}_{22}=0,\\
\label{e23}
&&l=1,2,3,4 , \quad -2\xi\frac{{\dot Z}_6}{Z_6}=Z_1-1.
\eeq
The last relation  is equivalent to (\ref{e6}).

Below, in section 5, we find that all constants  ``$Z$'' can be
interpreted as renormalization constants which are uniquely defined
from the conditions of reducing divergences.

Now we formulate the results obtained in this subsection as the following
lemma.

{\it Lemma}: Let
\beq
\label{23}
P=\int dx
P(Q,Q^*,\overline{C},B,{\cal B},\xi,\theta,\chi),
\eeq
be a local
functional of all variables with quantum numbers of the action
$S_{ext}$ and satisfies all equations  (\ref{a21}) - (\ref{a25}) and
additional symmetries (the axial- ,Poincare - invariance and so on)
which were used in solving the equations (\ref{a21}) - (\ref{a25})
(with substitution $S_{ext}\to P$ ). Then the functional $P$ has the
form
\beq
 \label{e24}
  &&\qquad\qquad\qquad
P=\Gamma_{0|0}+\sum_{k=6}^{22}Z_k\Gamma_{0|k}+
\chi\sum_{k=1}^5Z_k\Gamma_{0|k}, \\
\nonumber
&&Z_8=Z_9=Z_{10}=Z_{11}=Z_{12}=\frac{Z_7}{Z_6}, \; Z_{14}=\frac{Z_{13}}{Z_6},
\; Z_{15}=Z_{13}Z_{23}, \; Z_{17}=\frac{Z_{16}}{Z_6},  \\
\label{e24a}
&&Z_{18}=\frac{Z_{16}}{Z_6^2}, \; Z_{19}=Z_{16}Z_{24}, \; Z_{20}=Z_{16}^2Z_{25},
\; Z_{21}=Z_{13}Z_{16}^{1/2}Z_{26}, \\
\nonumber
&&Z_1=1-2\xi\frac{\dot{Z}_6}{Z_6}, \; Z_2=1+2\xi\Big(\frac{\dot{Z}_7}{Z_7}-
\frac{\dot{Z}_6}{Z_6}\Big), \; Z_4=2+2\xi\frac{\dot{Z}_{13}}{Z_{13}}-Z_3,\\
 \label{e24b}
&&Z_5=1+2\xi\frac{\dot{Z}_{16}}{Z_{16}},\quad\dot{Z}_{22}\!=\!\dot{Z}_{23}\!=
\!\dot{Z}_{24}\!=\!\dot{Z}_{25}\!=\!\dot{Z}_{26}\!=\!0,
\eeq
where $Z_6,Z_7,Z_{13},Z_{15},Z_{16},Z_{19},Z_{20},Z_{21},Z_{22}$  are arbitrary
constants depending perhaps on $\xi$, the constants   $Z_{22},Z_{23},Z_{24},
Z_{25}, Z_{26}$ do not depend on $\xi$.
Among the set of  indices of all constants $``Z''$, $\{k\}=1,...,26$ we will
highlight the three groups: the first is indices of independent constants,
\beq
\label{e24c}
\{k_{ind}\}=3,6,7,13,16,22,23,24,25,26;
\eeq
 the  second  is indices of  constants independent of $\xi$,
\beq
\label{e24d}
\{k_{con}\}=22,23,24,25,26\subset\{k_{ind}\};
\eeq and the third is indices of dependent constants,
\beq
\label{e24e}
\{k_{dep}\}=1,2,4,5,8,9,10,11,12,14,15,17,18,19,20,21.
\eeq

It should be noted that the functional $P$ does not contain vertices
additional to that from which the action $S_{ext}$ (\ref{a11}) is
built up. Namely, the obvious relation is valid, \beq \label{e25}
S_{ext}=\left.P\right|_{``Z''=1}. \eeq In its turn, the functional
$P$ (\ref{e24}) can be represented in the form analogous to
(\ref{a11}) for the action $S_{ext}$,
\beq
\nonumber &&P=\int dx
\Big\{-\frac{Z_{22}}{4}G_{\mu\nu}^\alpha(U)G_{\mu\nu}^\alpha(U)+
Z_{13}i\overline{\psi}\gamma^\mu
D_{\psi\mu}(U)\psi+\frac{Z_{16}}{2}\left(
D_{\varphi\mu}^{\alpha\beta}(U)\varphi^\beta\right)D_{\varphi\mu}^{\alpha
\gamma}(U)\varphi^\gamma- \\
\nonumber
&&-Z_{15}m\overline{\psi}\psi-Z_{19}\frac{M^2}{2}\varphi^2-Z_{20}\frac
{\lambda}{4}\varphi^4+Z_{21}i\vartheta\varphi^\alpha\overline{\psi}t^\alpha\psi
+Z_6\theta^\alpha_\mu[D^{\alpha\beta}_\mu(U)\overline{C}^\beta-
A^{\ast\alpha}_\mu]+ \\
\nonumber
&&+\frac{Z_7}{Z_6}Q^*\mathcal{T}_Q+\chi[Z_1(A_\mu^{\ast\alpha}-D_\mu^{\alpha
\beta}(\mathcal{B})\overline{C}^\beta)A_\mu^\alpha+Z_2C^{*\alpha}C^\alpha+
Z_3\psi^\ast\psi+Z_4\overline{\psi}^*\overline{\psi}+Z_5\varphi^{*\alpha}
\varphi^\alpha]+ \\
\label{e26}
&&+(\xi/2)B^\alpha B^\alpha+B^\alpha D^{\alpha\beta}_\mu({\cal B})A^\beta_\mu+
Z_7\overline{C}^{\alpha}D^{\alpha\beta}_{\mu}(\mathcal{B})D^{\beta\gamma}_{\mu}
(U)C^{\gamma}\Big\},
\eeq
\beq
\label{e27}
\mathcal{T}_{A_\mu^\alpha}=Z_6D_\mu^{\alpha\beta}(U)C^\beta, \quad
\mathcal{T}_{\psi,\overline{\psi},\varphi,C}=
\mathcal{R}_{\psi,\overline{\psi},\varphi,C}\; .
\eeq
Representation  (\ref{e26}) for the functional $P$ shows  that it can be
considered as (renormalized) action of the Yang-Mills theory, and from
representations  (\ref{e24}) and  (\ref{e25}) of functionals  $P$ and
$S_{ext}$ it follows that the renormalization is multiplicative.

The inverse (maybe trivial but important) statement is valid: if the
functional $P$ has the form  (\ref{e24}) and the relations
(\ref{e24a}), (\ref{e24b}) are satisfied then this functional
satisfies the equations (\ref{a22}) - (\ref{a25}).

\section{Effective action}
\noindent It is useful to define the generating functional of Green
functions with the help of the functional $P$ constructed in the
previous subsection, because it allows one to obtain a finite theory
from the beginning. In what follows we redefine  the functional $P$,
$P\equiv S_R$, and, respectively,  $P^{(k)}\equiv S_R^{(k)}$,
$\tilde{P}^{(k)}\equiv\tilde{S}_R^{(k)}, k=1,2$.

The generating functional of Green functions is given by the
following functional integral
\beq
\label{b0}
Z(J_{\Phi},L)=\int
d\Phi\exp\Big(\frac{i}{\eta}\big[S_R+J_{\Phi}\Phi\big]\Big)
=\exp\Big\{\frac{i}{\eta }W(J_{\Phi},L)\Big\},
\eeq
where $\eta$ is
the parameter of a loop expansion of expression in the exponent
(\ref{b0}),  $W(J_{\Phi},L)$ is the generating functional of
connected Green functions, $J_{\Phi}$ is the set of sources to
fields $\Phi$ and the notations $\Phi=\{Q,\overline{C},B\}$ and
$L=\{L^A\}=\{\mathcal{B},Q^{\ast},\xi,\theta, \chi\}$ are
introduced. We suppose also that all ``$Z$'' are functions of
 $\eta$, $``Z''=``Z(\eta)''$ expanding in the Taylor series,   $Z_k(0)=1,
 \dot{Z}_k=O(\eta),\; k=1,...,26$. In that case the functional $S_R$ becomes a
 function of  $\eta$,
\beq
\label{b1}
S_R=S_R(\eta)=\sum_{l=0}^{\infty}\eta^lS_{R,l},\quad
S_R^{[k]}=\sum_{l=0}^k\eta^lS_{R,l},
\eeq
and all functionals $S_{R,l}$ are linear combinations of the same set of
polynomials $\{\Gamma_{0|k}, \; k=0,1,..., 22\}$ with the help of which the
functionals  $S_{ext}$ and $P$ are presented.

The generating functional of vertex functions (effective action)
is defined with the help of the Legendre transformation,
\beq
\label{b2}
\Gamma(\Phi_{m|},L)=W(J_{\Phi},L)-J_{\Phi}\Phi_{m|},\quad  \Phi_{m|}=
\frac{\delta}{\delta J_{\Phi}}W(J_{\Phi},L),\
\eeq
with quantum numbers $\varepsilon(\Gamma)=0$, ${\rm gh}(\Gamma)=0$,
${\rm dim}(\Gamma)=0$,  $\varepsilon_f(\Gamma)=0$ and satisfies the relations
\beq
\label{b3}
\Gamma (\Phi_{m|},L)\frac{\overleftarrow{\delta }}{\delta
\Phi_{m|}}=-J_{\Phi}(\Phi_{m|},L),\quad \Gamma
(\Phi_{m|},L)\frac{\overleftarrow{\delta }}{\delta L^{A}
}=W(J_{\Phi},L)\frac{\overleftarrow{\delta }}{\delta
L^{A}}
\eeq
Functional averaging of equations  (\ref{a22}) - (\ref{a25}) with substitution
$S_{ext} \to S_R$ leads to the equations for the functional
$\Gamma=\Gamma(\Phi_{m|},L)$ copying the equations for $S_R$,
\beq
&&\int dx\Big( \Gamma \frac{\overleftarrow{\delta }}{\delta Q_{m|}}\frac{
\delta }{\delta Q^{\ast }}\Gamma -B_{m|}\frac{\delta }{\delta \overline{C}
_{m|}}\Gamma -\theta \frac{\delta }{\delta \mathcal{B}}\Gamma \Big) +2\chi
\xi \frac{\partial }{\partial \xi }\Gamma +  \notag \\
\label{b4}
&&+\chi \int dx\Big[ \Big(Q_{m|}\frac{\delta}{\delta Q_{m|}} -
Q^*\frac{\delta}{\delta Q^*}-\overline{
C}_{m|}\frac{\delta }{\delta \overline{C}_{m|}}-B_{m|}\frac{\delta }{\delta
B_{m|}}\Big) \Gamma \Big] =0,
\eeq
\beq
\label{b5}
\Gamma \overleftarrow{H_{m|}^{\alpha }}\omega _{\alpha }=0,
\eeq
where  $\overleftarrow{H_{m|}^{\alpha }}\omega _{\alpha }$ is given by
expression (\ref{a23}) with substitution  $\Phi$ $\rightarrow \Phi_{m|}$,
\beq
\label{b6}
&&\frac{\delta}{\delta B_{m|}^\alpha}\Gamma=D_\mu^{\alpha\beta}(\mathcal{B})
A_{m|\mu}^\beta+\xi B_{m|}^\alpha,  \\
\label{b7}
&&D_\mu^{\alpha\beta}(\mathcal{B})\frac{\delta}{\delta A_\mu^{*\beta}}\Gamma-
\frac{\delta}{\delta\overline{C}_{m|}^\alpha}\Gamma=-
g\varepsilon^{\alpha\beta\gamma}A_{m|\mu}^\beta\theta_\mu^\gamma.
\eeq

Represent the functional $\Gamma$ in the form
\beq
\label{b8}
\Gamma=\Gamma _{0|0}+\Gamma^{(1)}+\chi \Gamma^{(2)},
\eeq
 where the functionals  $\Gamma ^{(1)}$ and  $\Gamma ^{(2)}$ do not depend on
the  parameter  $\chi $. Thanks to  the structure chosen for the
functional
 (\ref{b8}), from the equations (\ref{b6}) and (\ref{b7}), it follows that the
 functionals  $\Gamma ^{(1)}$ and $\Gamma ^{(2)}$ do not depend on the fields
 $B^{\alpha}_{m|}$,
\beq
\label{b10}
 \frac{\delta }{\delta B_{m|}}\Gamma ^{(k)}=0,\
\Gamma ^{(k)}=\Gamma^{(k)}
(Q_{m|},\overline{C}_{m|},\mathcal{B},Q^{\ast},\xi ,\theta),\ k=1,2,
\eeq
and satisfy the equations
\beq
\label{b11}
\Big(D_{\mu}^{\alpha\beta}(\mathcal{B})\frac{\delta}{\delta
A_{\mu}^{*\beta}} -\frac{\delta }{\delta \overline{C}_{m|}^{\alpha
}}\Big) \Gamma ^{(k)}=0,\ k=1,2.
\eeq
In its turn, the equation
(\ref{b4}) splits into two ones, one of them is closed with respect
to $\Gamma^{(1)}$,
\beq
\label{b12}
\int dx\Big[ \Gamma ^{(1)}\frac{\overleftarrow{\delta }}{\delta Q_{m|}}%
\frac{\delta}{\delta Q^*}\Gamma ^{(1)}-g\theta _\mu^\alpha\varepsilon^
{\alpha\beta\gamma}\overline{C}_{m|}^\beta\frac{\delta}{\delta A_\mu^{*\gamma}}
\Gamma^{(1)}-\theta_\mu^\alpha\frac{\delta }{\delta\mathcal{B}_\mu^\alpha}
\Gamma^{(1)}\Big] =0,
\eeq
 and the second includes both functionals and describes their  dependence on
 the gauge parameter  $\xi$,
 \beq
&&\!\!\!\!\!\!2\xi\frac{\partial}{\partial\xi}\Gamma^{(1)}\!=\!\int\! dx\Big[
\Gamma^{(1)}\Big(\frac{\overleftarrow{\delta}}{\delta Q_{m|}}\frac{\delta}
{\delta Q^{\ast}}-\frac{\overleftarrow{\delta}}{\delta Q^{\ast}}\frac{\delta}
{\delta Q_{m|}}\Big)\Gamma^{(2)}\!-\!\Big(g\theta_\mu^\alpha\varepsilon^
{\alpha\beta\gamma}\overline{C}_{m|}^{\beta}\frac{\delta}{\delta A_{\mu }^
{\ast\gamma}}+\theta_{\mu}^{\alpha}\frac{\delta}{\delta\mathcal{B}_{\mu}^{\alpha}}
\Big)\Gamma^{(2)}\Big]+ \notag \\
\label{b13}
&&\qquad\qquad +\int dx\Big[\Big(\overline{C}_{m|}\frac{\delta}{\delta
\overline{C}_{m|}}-Q_{m|}\frac{\delta}{\delta Q_{m|}}+Q^*\frac{\delta}
{\delta Q^*}\Big) \Gamma ^{(1)}\Big] .
\eeq
 Now, the equation  (\ref{b5}) rewrites in the form of the two equations for the
 functionals $\Gamma ^{(1)}$ and $\Gamma ^{(2)}$,
\beq
\label{b14}
\Gamma^{(k)}\overleftarrow{h_{m|}^\alpha}\omega_\alpha=0, \ k=1,2,
\eeq
where
\beq
\nonumber
&&\overleftarrow{h_{m|}^\alpha}\omega_\alpha=\int dx \Big\{\Big[
\frac{\overleftarrow{\delta}}{\delta\mathcal{B}_\mu^\beta}D_\mu^{\beta\alpha}
(\mathcal{B})+g\varepsilon^{\beta\gamma\alpha}\Big(\frac
{\overleftarrow{\delta}}{\delta A_{m|\mu}^\beta}A_{m|\mu}^\gamma+%
\frac{\overleftarrow{\delta}}{\delta\varphi_{m|}^\beta}\varphi_{m|}^\gamma +
\frac{\overleftarrow{\delta}}{\delta C_{m|}^\beta}C_{m|}^\gamma\Big)+   \\
\nonumber
&&\qquad\qquad\quad+g\varepsilon^{\beta\gamma\alpha}\Big(\frac
{\overleftarrow{\delta}}{\delta\overline{C}_{m|}^\beta}\overline{C}_{m|}^\gamma
+\frac{\overleftarrow{\delta}}{\delta A_\mu^{*\beta}}A_\mu^{*\gamma}+
\frac{\overleftarrow{\delta}}{\delta C^{*\beta}}C^{*\gamma}+
\frac{\overleftarrow{\delta}}{\delta\varphi^{*\beta}}\varphi^{*\gamma}+
\frac{\overleftarrow{\delta}}{\delta\theta_\mu^\beta}\theta_\mu^\gamma\Big)- \\
\label{b15}
&&\qquad\qquad\quad-gt_{jk}^\alpha\Big(\frac{\overleftarrow{\delta}}{\delta
\psi_{m|j}}\psi_{m|k}+\frac{\overleftarrow{\delta}}{\delta\overline{\psi}_j^*}
\overline{\psi}_k^*\Big)+g\Big(\frac{\overleftarrow{\delta}}{\delta
\overline{\psi}_{m|j}}\overline{\psi}_{m|k}\!+\!\frac{\overleftarrow{\delta}}
{\delta\psi_j^*}\psi_k^*\Big)t_{kj}^\alpha\Big]\omega_\alpha\Big\}.
\eeq

Due to the equations  (\ref{b11}) it is useful to introduce new
variables $\mathcal{A}_\mu^{*\alpha}=\mathcal{A}_\mu^{*\alpha}(x)$,
$\mathcal{A}_{m|\mu}^{*\alpha}=\mathcal{A}_{m|\mu}^{*\alpha}(x)$,
\beq
\label{b16}
\mathcal{A}_{\mu}^{\ast\alpha}=A_{\mu}^{\ast\alpha}-D_{\mu}^{\alpha\beta}
(\mathcal{B})\overline{C}^{\beta},\quad\mathcal{A}_{m|\mu}^{\ast\alpha}=
A_{\mu}^{\ast\alpha}-D_{\mu}^{\alpha\beta}(\mathcal{B})
\overline{C}_{m|}^{\beta},
\eeq
and  to use the following agreement
for the sake of uniformity of further notations
\beq \label{b17}
\mathcal{A}_{\mu }^{\alpha }=A_{\mu }^{\alpha }.
\eeq

Also, let us introduce new functionals $\tilde{\Gamma}^{(k)}$ by the
rule
\beq
\label{b18}
\tilde{\Gamma}^{(k)}(\mathcal{B},\overline{C}_{m|},\mathcal{A}_{m|}^{\ast},
\Lambda_{m|})=\Gamma^{(k)}(\mathcal{B},\overline{C}_{m|},A^{\ast},
\Lambda_{m|})\big|_{A^{\ast}\to\mathcal{A}_{m|}^{\ast}+D_{\mu}^{\alpha\beta}
(\mathcal{B})\overline{C}_{m|}^{\beta}}, \eeq where the following
notations \beq \label{b19}
\Lambda=\{Q,\psi^{\ast},\overline{\psi}^{\ast},C^{\ast},\xi,\theta\},\quad
\Lambda_{m|}=\{Q_{m|},\psi^{\ast},\overline{\psi}^{\ast},C^{\ast},\xi,\theta\}
\eeq were used. Taking into account the definitions  (\ref{b16}) -
(\ref{b18})  we have \beq \label{b20}
&&\frac{\delta}{\delta A_\mu^{*\alpha}}\Gamma^{(k)}=\frac{\delta}%
{\delta \mathcal{A}_{m|\mu}^{*\alpha}}\tilde{\Gamma}^{(k)},   \\
\label{b21}
&&\frac{\delta }{\delta \overline{C}_{m|}^{\alpha }}\Gamma ^{(k)}=\frac{%
\delta }{\delta \overline{C}_{m|}^{\alpha }}\tilde{\Gamma}^{(k)}+D_{\mu
}^{\alpha \beta }(\mathcal{B})\frac{\delta }{\delta \mathcal{A}_{m|\mu
}^{\ast \beta }}\tilde{\Gamma}^{(k)},   \\
\label{b22}
&&\frac{\delta }{\delta \mathcal{B}_{\mu }^{\alpha }}\Gamma ^{(k)}=\frac{%
\delta }{\delta \mathcal{B}_{\mu }^{\alpha }}\tilde{\Gamma}^{(k)}-gf^{\alpha
\beta \gamma }\overline{C}_{m|}^{\beta }\frac{\delta }{\delta \mathcal{A}%
_{m|\mu }^{\ast \gamma }}\tilde{\Gamma}^{(k)},\quad k=1,2. \eeq
Then, from the equations  (\ref{b11}), (\ref{b18}),  (\ref{b20}),
(\ref{b21}), we find that \beq \label{b23} \frac{\delta }{\delta
\overline{C}_{m|}^{\alpha }}\tilde{\Gamma}^{(k)}=0,
\eeq
the
functionals $\tilde{\Gamma}^{(k)}, k=1,2$ do not depend on the
fields $\overline{C}_{m|}^{\alpha}$,
\beq
\label{b24}
\tilde{\Gamma}^{(k)}=\tilde{\Gamma}^{(k)}(\Omega_{m|},\Omega
_{m|}^*, \mathcal{B},\xi,\theta) .
\eeq
Here and below we will use
the notations
\beq
\label{b25}
\Omega_{m|}=\{\mathcal{A}_{m|},\psi_{m|},\overline{\psi}_{m|},\varphi_{m|},
C_{m|}\}, \quad \quad \Omega_{m|}^*=\{\mathcal{A}_{m|}^*,\psi^*,
\overline{\psi}^*,\varphi^*,C^*\} .
\eeq
Now, thanks to
(\ref{b18}), (\ref{b24}), (\ref{b25}), the equations (\ref{b12}) and
(\ref{b13}) rewrite as
\beq
\label{b26}
&&\qquad\qquad\qquad
\frac{1}{2}(\tilde{\Gamma}^{(1)},\tilde{\Gamma}^{(1)})-\int
dx\Big(\theta
\frac{\delta}{\delta\mathcal{B}}\Big)\tilde{\Gamma}^{(1)}=0, \\
&&\!\!\!\!2\xi\frac{\partial}{\partial\xi}\tilde{\Gamma}^{(1)}=(\tilde
{\Gamma}^{(1)},\tilde{\Gamma}^{(2)})+\int
dx\Big(\Omega_{m|}^*\frac{\delta}
{\delta\Omega_{m|}^*}-\Omega_{m|}\frac{\delta}{\delta\Omega_{m|}}\Big)
\tilde{\Gamma}^{(1)}\!-\!\int \!dx\Big(\theta\frac{\delta}{\delta
\mathcal{B}} \Big)\tilde{\Gamma}^{(2)}, \label{b27}
\eeq
where the
notation for antibracket \cite{BV,BV1}
\beq
\label{b27a}
&&(F,G)=\frac{1}{2}F\int dx\Big(\frac{\overleftarrow{\delta}}{\delta
\Omega_{m|}}\frac{\delta}{\delta\Omega_{m|}^{\ast}}-
\frac{\overleftarrow{\delta}}{\delta\Omega_{m|}^{\ast}}
\frac{\delta}{\delta\Omega_{m|}}\Big)G
\eeq
 is used. Further,
taking into account  (\ref{b18}), (\ref{b24}) and
\beq
\nonumber
&&\Gamma^{(k)}\int
dx\Big[\frac{\overleftarrow{\delta}}{\delta\mathcal{B}_\mu^
\beta}D_\mu^{\beta\alpha}(\mathcal{B})+g\varepsilon^{\beta\gamma\alpha}
\Big( \frac{\overleftarrow{\delta }}{\delta \overline{C}_{m|}^\beta}
\overline{C}_{m|}^\gamma+\frac{\overleftarrow{\delta}}{\delta
A_\mu^{*\beta}}
A_\mu^{*\gamma}\Big)\Big] = \\
\label{b29} &&=\tilde{\Gamma}^{(k)}\int
dx\Big[\frac{\overleftarrow{\delta}}{\delta
\mathcal{B}_\mu^\beta}D_\mu^{\beta\alpha}(\mathcal{B})+g\varepsilon^
{\beta\gamma\alpha}\frac{\overleftarrow{\delta}}{\delta\mathcal{A}_{m|\mu}^
{*\beta}}\mathcal{A}_{m|\mu}^{*\gamma}\Big], \eeq we find that \beq
\label{b30}
\Gamma^{(k)}\overleftarrow{h_{m|}^{\alpha}}\omega_{\alpha}=\tilde{\Gamma}^{(k)}
\overleftarrow{\tilde{h}_{m|}^{\alpha}}\omega_{\alpha}=0,\ k=1,2,
\eeq
where the operator
$\overleftarrow{\tilde{h}_{m|}^{\alpha}}\omega_{\alpha}$ is defined
in the equality  (\ref{c14}) with the substitution $\Omega
\rightarrow \Omega_{m|}, \;\Omega^*\rightarrow \Omega^*_{m|}$.

Later, on when studying the tensor structure of divergent parts of
the generating functional of vertex functions, it is useful to use
the consequence of the equation (\ref{b30}) corresponding to the
particular case of $T_{m|}$-symmetry when  $\omega _{\alpha
}(x)=\mathrm{const}$,
\beq
\label{b32}
\tilde{\Gamma}^{(k)}\overleftarrow{T_{m|}^{\alpha }}=0,\ k=1,2,
 \eeq
where the operators  $\overleftarrow{T_{m|}^{\alpha }}$ are defined
by the equalities (\ref{c16}) with substitution $\Omega \rightarrow
\Omega_{m|}, \; \Omega^*\rightarrow \Omega^*_{m|}$.

\section{Multiplicative renormalization}
\noindent In this section, we study a structure of renormalization
in the model under consideration, and find multiplicative character
of renormalizability. The main role is played by solving the
extended master-equation  (\ref{c11}) and the equation describing
the gauge dependence  (\ref{c19}). We will show that the
renormalized quantum action and the effective action satisfy  this
equation exactly to every order of loop expansions. The structure of
the renormalized action is determined by the same monomials in
fields and antifields as the non-renormalized quantum action does
but with constants defined by the divergences of the effective
action. For simplicity of notations, we will often omit the lower
index  $m|$ in arguments of functional $\Gamma$.

\subsection{Tree approximation ($\eta =0$)}
\noindent
Consider the tree approximation for the functional $\Gamma$,
$\Gamma_{0}=S_{\mathrm{ext}}$, which in new variables reads
\begin{eqnarray}
\label{f1}
&&\Gamma_{0}=\Gamma _{0|0}+\Gamma_{0}^{(1)}+\chi\Gamma_{0}^{(2)}, \\
\label{f2}
&&\Gamma _{0}^{(1)}=\tilde{\Gamma}_{0}^{(1)},\quad
\Gamma_{0}^{(2)}=\tilde{\Gamma}_{0}^{(2)},
\end{eqnarray}
Represent the functional  $\tilde\Gamma^{(1)}_0$ in the form
\beq
\label{da18a}
&&\tilde\Gamma^{(1)}_0=\Gamma_{0\theta}+
\Gamma_{0\Omega^*}+\Gamma_{0\psi} +\Gamma_{0\varphi}+
\Gamma_{0\psi\varphi}+\Gamma_{0\mathcal{AB}},  \\
\label{da18b}
&&\Gamma_{0\Omega^*}=\Gamma_{0\mathcal{A}^*}+\Gamma_{0C^*}+
\Gamma_{0\psi^*}+\Gamma_{0\overline{\psi}^*}+\Gamma_{0\varphi^*},
\eeq
where the following notations useful for further calculations will be used,
\beq
\label{da19}
&&\Gamma_{0\theta}=\Gamma_{0|6}, \quad \Gamma_{0\mathcal{A}^*}=\Gamma_{0|7}+
\Gamma_{0|8}, \quad \Gamma_{0C^*}=\Gamma_{0|9}, \\
\label{da21}
&&\Gamma_{0\psi^*}=\Gamma_{0|10}, \quad \Gamma_{0\overline{\psi}^*}=
\Gamma_{0|11}, \quad \Gamma_{0\varphi^*}=\Gamma_{0|12}, \\
\label{da22}
&&\Gamma_{0\psi}=\Gamma_{0|13}+\Gamma_{0|14}+\Gamma_{0|15}, \\
\label{da23}
&&\Gamma_{0\varphi}=\Gamma_{0|16}+\Gamma_{0|17}+\Gamma_{0|18}+
\Gamma_{0|19}+\Gamma_{0|20}, \quad \Gamma_{0\psi\varphi}=\Gamma_{0|21}, \\
\label{da24}
&&\Gamma_{0\mathcal{AB}}=\Gamma_{0|22}.
\eeq

The functional $\tilde{\Gamma}_0^{(2)}$  has the form
\beq
\label{da25}
\tilde{\Gamma}_0^{(2)}=\Gamma_{0|1}+\Gamma_{0|2}+\Gamma_{0|3}+
+\Gamma_{0|4}+\Gamma_{0|5}.
\eeq

Remind that the functional $\Gamma_0$ satisfies the equation
(\ref{a22}) - (\ref{a25}).

\subsection{(l+1)-loop approximation (order $\eta^{l+1}$)}
\noindent The proof of multiplicative renormalizability will be
given by the method of mathematical induction  in loop expansions of
the effective action applying  the scheme of minimal subtractions.
To this end, let us suppose  we have found the parameters
$Z_{k_{ind}}$,
\beq
\nonumber
&&Z_{k_{ind}}=Z_{k_{ind}}^{[l]}+O(\eta^{l+1}), \quad
Z_{k_{ind}}^{[l]}=1+
\sum_{n=1}^l\eta^n z_{k_{ind},n}, \; \forall k_{ind}, \\
\label{ca1} &&\dot{z}_{k_{con},n}=0, \;\;  \forall k_{con},\;\;
1\leq n\leq l,
\eeq
so that the $l$-loop approximation for $\Gamma$,
$\Gamma^{[l]}=\sum_{n=0}^l \eta^n\Gamma_n$, is a finite functional.
We will show that the $(l+1)$-loop approximation for $Z_{k_{ind}}$
can be picked up so that,
\beq
\label{ca1a1}
Z_{k_{ind}}=Z_{k_{ind}}^{[l+1]}+O(\eta^{l+2}), \quad
Z_{k_{ind}}^{[l+1]}= Z_{k_{ind}}^{[l]}+z_{{k_{ind}},l+1},
\eeq
which compensates divergences of the $(l+1)$-loop approximation of the
functional $\Gamma$.

Represent the action $S_R$ in the form
\beq
\label{ca2}
S_R=S_R^{[l]}+\eta^{l+1}s_{l+1}+O(\eta^{l+2}),
\eeq
where  $S_R^{[l]}$ is the action $S_R$ with independent parameters
$Z_{k_{ind}}$ replaced by $Z_{k_{ind}}^{[l]}$ satisfying the equations
(\ref{a22})-(\ref{a25}) and functional  $s_{l+1}$ is equal to
\beq
\label{ca3}
s_{l+1}=s_{l+1}^{(1)}+\chi s_{l+1}^{(2)}.
\eeq

In this subsection we will use the short notations for variational
derivatives of the type
\beq
\label{ca4}
\frac{\delta}{\delta\mathcal{A}}\to\pa_{\mathcal{A}},\quad
\mathcal{A}\pa_{\mathcal{A}}=\int dx
\mathcal{A}\frac{\delta}{\delta\mathcal{A}},
\eeq
when it will not lead to uncertainties.

Now let us study the structure of the functional $\Gamma$ with
taking into account the  $(l+1)$-loop approximation. It describes by
diagrams with  vertices of the action $S_R$ with parameters
$z_{k_{ind},n}$, $0\leq n\leq l+1$, $z_{k_{ind},0}=0$, i.e. with
vertices of action $S_R^{[l]}$ and with vertices of  $s_{l+1}$.
Because we are interested in diagrams of loop order not higher than
$l+1$, the vertices from $s_{l+1}$ cannot appear in loop diagrams,
i.e.  the vertices from $s_{l+1}$ give "tree" contribution to
$\Gamma$ equal to  $\eta^{l+1}s_{l+1}$. The rest diagrams are
generated by the action $S_R^{[l]}$. Denote the contribution of
these diagrams to functional $\Gamma$ as $\Gamma(S_R^{[l]})$, i.e.,
\beq
\label{ca5}
\Gamma=\Gamma(S_R^{[l]})+\eta^{l+1}s_{l+1}+O(\eta^{l+2}).
\eeq
Because the action $S_R^{[l]}$ satisfies the equations  (\ref{a22})
- (\ref{a25}), the functional $\Gamma(S_R^{[l]})$ satisfies these
equations too (with substitutions $Q,\overline{C},B\rightarrow
Q_{m|},\overline{C}_{m|},B_{m|}$).

Represent the functional $\Gamma(S_R^{[l]})$ in the form
\beq
\label{ca6}
\Gamma(S_R^{[l]})=\Gamma_{0|0}+\Gamma^{(1)}(S_R^{[l]})+
\chi\Gamma^{(2)}(S_R^{[l]}),
\eeq
Repeating calculations made in section 3 we find that
\beq
\label{ca7}
\Gamma^{(k)}(S_R^{[l]}|Q_{m|},\overline{C}_{m|},\mathcal{B},Q^{\ast},B_{m|},
\xi,\theta)=\tilde{\Gamma}^{(k)}(S_R^{[l]}|\Omega_{m|},\Omega_{m|}^{\ast},
\mathcal{B},\xi,\theta),\ k=1,2,
\eeq
and functionals $\tilde{\Gamma}^{(k)}(S_R^{[l]})$ satisfy the equations
(\ref{b26}), (\ref{b27}) and (\ref{b30}).

Represent the functionals $\tilde{\Gamma}^{(k)}(S_R^{[l]})$ in the
sum of divergent and finite (after removing a regularization) terms.
Taking into account that the functionals
$\tilde{\Gamma}^{(k)}(S_R^{[l]})$ are finite to $n$-loop
approximations, $0\leq n\leq l$, by proposition, we obtain
\beq
\label{ca8}
\tilde{\Gamma}^{(k)}(S_R^{[l]})=\tilde{\Gamma}^{(k)}(S_R^{[l]})_{\mathrm{fin}}
+\eta^{l+1}\tilde{\Gamma}^{(k)}(S_R^{[l]})_{l+1,\mathrm{div}}+O(\eta^{l+2}).
\eeq
To simplify the presentations,  we introduce the notations
\beq
\label{ca8a}
\tilde{\Gamma}^{(k)}(S_R^{[l]})_{l+1,\mathrm{div}}=u_{l+1}^{(k)},\;
u_{l+1}^{(1)}+\chi u_{l+1}^{(2)}=u_{l+1}, \;  k=1,2. \eeq Then, \beq
\label{ca9}
\Gamma=\Gamma(S_R^{[l]})_{\mathrm{fin}}+\eta^{l+1}\left(u_{l+1}+
s_{l+1}\right)+O(\eta^{l+2}),
\eeq
 and functionals $u_{l+1}^{(k)}$
are local polynomials of their arguments with quantum numbers of the
action $S_{ext}$ and contain only divergent  summands (formalism of
minimal subtraction). Then, they satisfy the equations which follow
from the equations (\ref{b26}), (\ref{b27}) and (\ref{b30}),
 \beq
\label{ca10}
&&\Big(\tilde{\Gamma}_0^{(1)},u_{l+1}^{(1)}\Big)-\int dx\Big(\theta
\frac{\delta}{\delta\mathcal{B}}\Big)u_{l+1}^{(1)}=0, \\
\nonumber
&&2\xi\frac{\partial}{\partial\xi}u_{l+1}^{(1)}=(\tilde{\Gamma}_0^{(1)},
u_{l+1}^{(2)})-(\tilde{\Gamma}_0^{(2)},u_{l+1}^{(1)})- \\
\label{ca11}
&&-\int dx\Big(\theta\frac{\delta}{\delta\mathcal{B}}\Big)u_{l+1}^{(2)}+
\int dx\Big(\Omega_{m|}^*\frac{\delta}{\delta\Omega_{m|}^*}-\Omega_{m|}
\frac{\delta}{\delta\Omega_{m|}}\Big)u_{l+1}^{(1)}, \\
\label{ca12}
&&u_{l+1}^{(k)}\overleftarrow{h^\alpha}\omega_\alpha=0,  \\
\label{ca13}
&&u_{l+1}^{(k)}\overleftarrow{T^\alpha}=0,
\quad k=1,2.
\eeq
Notice that the form of equations  (\ref{ca10}) - (\ref{ca13}) does not depend
on index $l$.

Taking into account the quantum numbers, the axial-, Poincare- and
T-symmetries we find general expression for local functional
$u_{l+1}^{(2)}$,
\beq
\label{ca14}
u_{l+1}^{(2)}=\sum_{k=1}^5q_{k,l+1}\Gamma_{0|k}+
q_{1,l+1}^{\prime}\mathcal{A}^*\mathcal{B},
\eeq
where  $q_{k,l+1},
k=1,...,5,$ and $q_{1,l+1}^{\prime}$ are arbitrary constants.
Further, using the equations  (\ref{ca12}) for  $u_{l+1}^{(2)}$, we
find $q_{1,l+1}^{\prime}=0$. Final expression for
$q_{1,l+1}^{\prime}=0$ reads
\beq
\label{ca15}
u_{l+1}^{(2)}=\sum_{k=1}^5q_{k,l+1}\Gamma_{0|k}.
\eeq
Notice that
the functional $u_{l+1}^{(2)}$ does not depend on the fields
$\theta$ and $\mathcal{B}$.

Due to the expression (\ref{ca15}) for  $u_{l+1}^{(2)}$, the equation
(\ref{ca11}) reduces to
\beq
\label{ca16}
2\xi\frac{\pa}{\pa\xi}u_{l+1}^{(1)}=\hat{L}(1+q_{,l+1})\tilde{\Gamma}_{0}^{(1)},
\eeq
where the operator $\hat{L}$ is defined in (\ref{c19}) with substitution
$Z_k \rightarrow 1+q_{k,l+1}, \;k=1,...,5.$

\subsubsection{Solution to equation (\ref{ca10})}
\noindent
Consider a solution to the equation  (\ref{ca10}) for the functional
$\tilde\Gamma^{(1)}_{l+1,\mathrm{div}}=\tilde{\Gamma}(S_R^{[l]})_{l+1,
\mathrm{div}}^{(1)}=u_{l+1}^{(1)}$ using the presentation
\beq
\nonumber
&&u_{l+1}^{(1)}=M_{\theta,l+1} +M_{\Omega^*,l+1}+M_{\psi,l+1}+M_{\varphi,l+1}+
M_{\psi\varphi,l+1}+M_{\mathcal{AB},l+1}(\mathcal{A,B}), \\
\label{da1}
&&M_{\Omega^*,l+1}=M_{\mathcal{A}^*,l+1}+M_{C^*,l+1}+M_{\psi^*,l+1}+
M_{\overline{\psi}^*,l+1}+M_{\varphi^*,l+1}.
\eeq
To this end, we
will find the general form of the functional
$\tilde\Gamma^{(1)}_{l+1,\mathrm{div}}$, following from locality,
quantum numbers, axial-, Poincare-, $T$- symmetries and, partially,
of the  gauge symmetry with respect to the external field ${\cal
B}$. In fact, all needed calculations copy calculations made in
section 3 when constructing the general form of the functional
$\tilde{P}^{(1)}$ (see formulas  (\ref{d2}) - (\ref{d16})) with
evident substitutions like $\tilde{P}_{\theta}^{(1)}\to M_{\theta}$.
Here we give the final results only. The functional $M_{\theta,l+1}$
reads \beq \label{daa3} M_{\theta,l+1}=q_{6,l+1}\Gamma_{0|6}. \eeq
For the functionals linear in antifields we find
\beq
\label{da5}
&&M_{\mathcal{A}^*,l+1}=q_{7,l+1}\Gamma_{0|7}+q_{8,l+1}\Gamma_{0|8}, \\
\label{da6}
&&M_{C^*,l+1}=q_{9,l+1}\Gamma_{0|9},  \\
\label{da7} &&M_{\psi^*,l+1}=q_{10,l+1}\Gamma_{0|10},  \;
M_{\overline{\psi}^*,l+1}=q_{11,l+1}\Gamma_{0|11}, \;
M_{\varphi^*,l+1}=q_{12,l+1}\Gamma_{0|12}. \eeq For the functionals
$M_{\psi,l+1}$, $M_{\varphi,l+1}$, $M_{\psi\varphi,l+1}$, we obtain
\beq \label{da16}
&&M_{\psi,l+1}=q_{13,l+1}\Gamma_{0|13}+q_{14,l+1}\Gamma_{0|14}+
q_{15,l+1}\Gamma_{0|15}, \\
&&M_{\varphi,l+1}=q_{16,l+1}\Gamma_{0|16}+q_{17,l+1}\Gamma_{0|17}+
M_{\varphi18,l+1}+q_{19,l+1}\Gamma_{0|19}+q_{20,l+1}\Gamma_{0|20}, \\
\label{da17}
&&M_{\psi\varphi,l+1}=q_{21,l+1}\Gamma_{0|21}, \\
\label{da18} &&M_{\varphi18,l+1}=\!\int
dx\!\left[\frac{g^2}{2}q_{18}^
{\alpha\beta\gamma\sigma}\mathcal{A}_\mu^\alpha\varphi^\beta
\mathcal{A}_\mu^\gamma\varphi^\sigma\right], \;
q_{18}^{\alpha\beta\gamma\sigma}=q_{18}^{\gamma\beta\alpha\sigma}=
q_{18}^{\alpha\sigma\gamma\beta}.
\eeq
Consider the consequences of
the equality to zero for contributions proportional to $\theta$.
\beq
\nonumber &&\theta\overline{\psi}\psi \; \Rightarrow
(q_{6,l+1}+q_{14,l+1})\pa_
{\mathcal{A}_\mu^\alpha}\Gamma_{0|14}-q_{13,l+1}\pa_{\mathcal{B}_\mu^\alpha}
\Gamma_{0|13}=0 \Rightarrow \\
\label{da19a}
&&\qquad\qquad\quad  q_{14,l+1}=q_{13,l+1}-q_{6,l+1};\\
\nonumber
&&\theta\mathcal{A}^*C \; \Rightarrow (q_{6,l+1}+q_{8,l+1})\pa_{\mathcal{A}_
\mu^\alpha}\Gamma_{0|8}-q_{7,l+1}\pa_{\mathcal{B}_\mu^\alpha}\Gamma_{0|7}=0
\; \Rightarrow \\
\label{da20}
&&\qquad\qquad\quad q_{8,l+1}=q_{7,l+1}-q_{6,l+1};\\
\nonumber
&&\theta\Big(D_\varphi(\mathcal{B})\varphi\Big)\varphi \; \Rightarrow
(q_{6,l+1}+q_{17,l+1})\pa_{\mathcal{A}_\mu^\alpha}\Gamma_{0|17}-
q_{16,l+1}\pa_{\mathcal{B}_\mu^\alpha}\Gamma_{0|16}=0 \Rightarrow \\
\label{da21a}
&&\qquad\qquad\quad q_{17,l+1}=q_{16,l+1}-q_{6,l+1};\\
\nonumber
&&\theta\mathcal{A}\varphi\varphi \; \Rightarrow q_{6,l+1}\pa_{\mathcal{A}_
\mu^\alpha}\Gamma_{0|18}+\pa_{\mathcal{A}_\mu^\alpha}M_{\varphi18,l+1}-
q_{17,l+1}\pa_{\mathcal{B}_\mu^\alpha}\Gamma_{0|17}=0 \Rightarrow \\
\label{da22a}
&&\qquad\qquad\quad M_{\varphi18,l+1}=q_{18,l+1}\Gamma_{0|18}, \quad
q_{18,l+1}=q_{16,l+1}-2q_{6,l+1};\\
\nonumber
&&\theta\mathcal{A}^n\mathcal{B}^m \; \Rightarrow q_{6,l+1}
\pa_{\mathcal{A}_\mu^\alpha}\Gamma_{0|22}(V)+
\pa_{\mathcal{A}_\mu^\alpha}M_{\mathcal{AB},l+1}-
\pa_{\mathcal{B}_\mu^\alpha}M_{\mathcal{AB},l+1}=0
\; \Rightarrow \\
\label{da23a}
&&\qquad\qquad\quad M_{\mathcal{AB},l+1}=-q_{6,l+1}\mathcal{A}
\pa_{\mathcal{A}}\Gamma_{0|22}(V)+M_{\mathcal{AB},l+1}^\prime(V),
\eeq
where $V=\mathcal{A}+\mathcal{B}$ and $M'_{\mathcal{AB},l+1}(V)$ is an
arbitrary functional of $V$.

When  $\theta=0$, the equation (\ref{ca10}) reduces to
\beq
\nonumber
&&\int
dx(\Gamma_{0\Omega^*}+\Gamma_{0\psi}+\Gamma_{0\varphi}+
\Gamma_{0\psi\varphi}+\Gamma_{0\mathcal{AB}})
\left(\overleftarrow{\pa}_{\Omega}\pa_{\Omega^*}-
\overleftarrow{\pa}_{\Omega^*}\pa_{\Omega}\right) \times\\
\label{dab24}
&&\times (M_{\Omega^*,l+1}+M_{\psi,l+1}+M_{\varphi,l+1}+
M_{\psi\varphi,l+1}+M_{\mathcal{AB},l+1}=0,
\eeq
which is not more than linear in antifields.

We have
\beq
\nonumber
&&\mathcal{A}^*\mathcal{A}CC \; \Rightarrow \; 2q_{8,l+1}\Gamma_{0|8}\frac
{\overleftarrow{\delta}}{\delta\mathcal{A}_\mu^\alpha}\frac{\delta}
{\delta\mathcal{A}_\mu^{*\alpha}}\Gamma_{0|8}+(q_{8,l+1}+q_{9,l+1})
\Gamma_{0|8}\frac{\overleftarrow{\delta}}{\delta C^\alpha}\frac{\delta}
{\delta C^{*\alpha}}\Gamma_{0|9}=0 \; \Rightarrow \\
\label{dab27}
&&\qquad\qquad\qquad  q_{9,l+1}=q_{8,l+1}=q_{7,l+1}-q_{6,l+1}; \\
\nonumber
&&\psi^*\psi CC \; \Rightarrow \; (q_{9,l+1}+q_{10,l+1})\Gamma_{0|10}\frac{
\overleftarrow{\delta}}{\delta C^\alpha}\frac{\delta}{\delta C^{*\alpha}}
\Gamma_{0|9}+2 q_{10,l+1}\Gamma_{0|10}\frac{\overleftarrow{\delta}}
{\delta\psi_j}\frac{\delta}{\delta\psi_j^*}\Gamma_{0|10}=0 \; \Rightarrow \\
\label{dab28}
&&\qquad\qquad\qquad  q_{10,l+1}=q_{9,l+1}=q_{7,l+1}-q_{6,l+1}; \\
\nonumber
&&\overline{\psi}\overline{\psi}^* CC \; \Rightarrow \; (q_{9,l+1}+q_{11,l+1})
\Gamma_{0|11}\frac{\overleftarrow{\delta}}{\delta C^\alpha}\frac{\delta}
{\delta C^{*\alpha}}\Gamma_{0|9}+2 q_{11,l+1}\Gamma_{0|11}
\frac{\overleftarrow{\delta}}{\delta\overline{\psi}_j}\frac{\delta}
{\delta\overline{\psi}_j^*}\Gamma_{0|10}=0 \; \Rightarrow \\
\label{dab29}
&&\qquad\qquad\qquad  q_{11,l+1}=q_{9,l+1}=q_{7,l+1}-q_{6,l+1};\\
\nonumber
&&\overline{\psi}\psi\varphi C \; \Rightarrow \; (q_{12,l+1}+q_{21,l+1})
\Gamma_{0|21}\frac{\overleftarrow{\delta}}{\delta\varphi^\alpha}\frac{\delta}
{\delta\varphi^{*\alpha}}\Gamma_{0|12}+(q_{10,l+1}+q_{21,l+1})\Gamma_{0|21}
\frac{\overleftarrow{\delta}}{\delta\psi_j}\frac{\delta}{\delta\psi_j^*}
\Gamma_{0|10}+ \\
\nonumber
&&\qquad\qquad\qquad  +(q_{11,l+1}+q_{21,l+1})\Gamma_{0|21}\frac
{\overleftarrow{\delta}}{\delta\overline{\psi}_j}\frac{\delta}
{\delta\overline{\psi}_j^*}\Gamma_{0|11}=0 \; \Rightarrow \\
\label{dab30}
&&\qquad\qquad\qquad  q_{12,l+1}=q_{10,l+1} =q_{7,l+1}-q_{6,l+1},
\eeq
where the relation $q_{11,l+1}=q_{10,l+1}$ was used.
\beq
\nonumber
&&\!\!\mathcal{A}^n\mathcal{B}^kC \; \Rightarrow \Gamma_{0|22}\frac{
\overleftarrow{\delta}}{\delta\mathcal{A}_\mu^\alpha}\frac{\delta}
{\delta\mathcal{A}_\mu^{*\alpha}}(q_{7,l+1}\Gamma_{0|7}+q_{8,l+1}\Gamma_{0|8})
-(\Gamma_{0|7}+\Gamma_{0|8})\frac{\overleftarrow{\delta}}{\delta\mathcal{A}_
\mu^{*\alpha}}\frac{\delta}{\delta\mathcal{A}_\mu^\alpha}
M_{\mathcal{AB},l+1}=0 \; \Rightarrow \\
\label{dab31}
&&\!\!M_{\mathcal{AB},l+1}^\prime=q_{22,l+1}\Gamma_{0|22}(V), \quad
M_{\mathcal{AB},l+1}=-q_{6,l+1}\mathcal{A}
\pa_{\mathcal{A}}\Gamma_{0|22}(V)+q_{22,l+1}\Gamma_{0|22}(V). \eeq
Thus, the divergences $u_{l+1}^{(1)}$ can be represented in the form
\beq \label{dab32}
u_{l+1}^{(1)}=\sum_{k=6}^{21}q_{k,l+1}\Gamma_{0|k}+M_{\mathcal{AB},l+1},
\eeq
and the functional $M_{\mathcal{AB},l+1}$ is given by the
formula (\ref{dab31}). As it was noted above, all functionals
considered as polynomials are independent,  and polynomials
$\Gamma_{0|6} - \Gamma_{0|21}$ are homogeneous with respect to the
fields $\Omega$ and the antifields $\Omega^*$.

\subsubsection{Solution to  equation (\ref{ca16})}
\noindent Taking into consideration that the polynomials
$\Gamma_{0|k}, k=6,...,21$, are eigen for operator $\hat{L}$,  we
find that eq. (\ref{ca16}) is reduced to the set of equations
\beq
\label{5.2.2.3}
&&2\xi\frac{\pa}{\pa\xi}q_{k,l+1}=\varrho_k, \; k=6,...,21, \\
\label{5.2.2.4}
&&2\xi\frac{\pa}{\pa\xi}M_{\mathcal{AB},l+1}=q_{1,l+1}\mathcal{A}
\pa_{\mathcal{A}}\Gamma_{0|22}(V),
\eeq
where $\varrho_k$ are the eigenvalues of the operator $\hat{L}$.

We have
\beq
\label{5.2.2.5}
&&k=6 \; \Rightarrow 2\xi\dot{q}_{6,l+1}=-q_{1,l+1} \quad
\Rightarrow \; q_{1,l+1}=-2\xi\dot{q}_{6,l+1};\\
\label{5.2.2.6}
&&k=7 \; \Rightarrow \; 2\xi\dot{q}_{7,l+1}=-q_{1,l+1}+q_{2,l+1} \;\;
\Rightarrow \; q_{2,l+1}=2\xi\dot{q}_{7,l+1}-2\xi\dot{q}_{6,l+1};\\
\label{5.2.2.7}
&&k=13 \; \Rightarrow \; 2\xi\dot{q}_{13,l+1}=q_{3,l+1}+q_{4,l+1} \;\;
\Rightarrow \; q_{4,l+1}=2\xi\dot{q}_{13,l+1}-q_{3,l+1};\\
\label{5.2.2.8}
&&k=15 \; \Rightarrow \; 2\xi\dot{q}_{15,l+1}=q_{3,l+1}+q_{4,l+1} \;\;
\Rightarrow \; \dot{q}_{23,l+1}=0, \;q_{15,l+1}=q_{13,l+1}\!+q_{23,l+1};\\
\label{5.2.2.9}
&&k=16 \; \Rightarrow \; 2\xi\dot{q}_{16,l+1}=2q_{5,l+1} \quad
\Rightarrow \; q_{5,l+1}=\xi\dot{q}_{16,l+1};\\
\label{5.2.2.10}
&&k=19 \; \Rightarrow \; 2\xi\dot{q}_{19,l+1}=2q_{5,l+1} \quad
\Rightarrow \; \dot{q}_{24,l+1}=0, \; q_{19,l+1}=q_{16,l+1}+q_{24,l+1};\\
\label{5.2.2.11}
&&k=20 \; \Rightarrow \; 2\xi\dot{q}_{20,l+1}=4q_{5,l+1} \quad
\Rightarrow \; \dot{q}_{25,l+1}=0, \; q_{20,l+1}=2q_{16,l+1}+q_{25,l+1};\\
\nonumber
&&k=21 \; \Rightarrow \; 2\xi\dot{q}_{21,l+1}=q_{3,l+1}+q_{4,l+1}+q_{5,l+1} \; \Rightarrow \\
\label{5.2.2.12} &&\qquad\qquad\qquad \dot{q}_{26,l+1}=0, \;
q_{21,l+1}=q_{13,l+1}+\frac{1}{2}q_{16,l+1}+q_{26,l+1}; \eeq
Finally, from (\ref{5.2.2.4}) it follows \beq \nonumber &&
2\xi\dot{q}_{22,l+1}\Gamma_{0|22}-
(2\xi\dot{q}_{6,l+1}+q_{1,l+1})\mathcal{A}\pa_{\mathcal{A}}\Gamma_{0|22}(V)
=0 \; \Rightarrow \\
\label{5.2.2.13}
&&2\xi\dot{q}_{22,l+1}\Gamma_{0|22}=0 \; \Rightarrow \dot{q}_{22,l+1}=0,
\eeq
where we used the result of the block $k=6$.

Finally, the solution to the set of eqs. (\ref{ca10}) - (\ref{ca13})
for $u_{l+1}^{(k)}, k=1,2$, can be represented in the form
\beq
\label{5.2.2.14}
&&u_{l+1}^{(1)}=\sum_{k=6}^{21}q_{k,l+1}\Gamma_{0|k}-q_{6,l+1}\mathcal{A}
\pa_{\mathcal{A}}\Gamma_{0|22}(V)+q_{22,l+1}\Gamma_{0|22}(V), \\
\label{5.2.2.15}
&&u_{l+1}^{(2)}=\sum_{k=1}^5q_{k,l+1}\Gamma_{0|k}.
\eeq
Let $\{k_{ind}\}=\{\{K_a\},22\}, \; \{K_a\}=\{3,6,7,13,16,23,...,26\}$.
Then we have
\beq
\label{5.2.2.16}
q_{k_{dep},l+1}=\sum_a\big([X_{q|k_{dep},a}+Y_{q|k_{dep},a}\xi\pa_\xi]q_{K_a,l+1}
\big),
\eeq
where the set $\{q_{k_{dep},l+1}\}=\{\{q_{K_a,l+1}\},q_{22}\}$ is the set of
arbitrary numbers, 5 of them (with indexes $3,6,7,13,16$) can depend on $\xi$
and $\dot{q}_{22,l+1}=\dot{q}_{23,l+1}=\dot{q}_{24,l+1}=\dot{q}_{25,l+1}=
\dot{q}_{26,l+1}=0$. Numerical matrices $X_{q|k_{dep},a}$ and $Y_{q|k_{dep},a}$
in eq. (\ref{5.2.2.16}) can be restored by using the relations
 (\ref{da19a}) - (\ref{da22a}),  (\ref{dab27}) - (\ref{dab30}),  (\ref{5.2.2.5}) - (\ref{5.2.2.12}).

\subsection{Finiteness  of effective action}
\noindent
Now, let us prove that one can choose the renormalization constants in  such a
way as to make  the effective action finite to the $(l+1)$-loop approximation.
To this end, we consider the divergent part of the effective action $\Gamma$,
$\Gamma_{l+1,\mathrm{div}}$, described by the equation (\ref{ca9}),
\beq
\label{e6a}
\Gamma_{l+1,\mathrm{div}}=u_{l+1}+s_{l+1}=u_{l+1}^{(1)}+s_{l+1}^{(1)}+
\chi[u_{l+1}^{(2)}+s_{l+1}^{(2)}].
\eeq
It is easy to see that the functional $s_{l+1}$ satisfies the same equations
(\ref{ca10}) - (\ref{ca13}) as $u_{l+1}$ does (with substitution $u_{l+1}^{(k)}
\; \rightarrow \; s_{l+1}^{(k)}, k=1,2$), such that we have
\beq
\label{5.2.2.14a}
&&s_{l+1}^{(1)}=\sum_{k=6}^{21}z_{k,l+1}\Gamma_{0|k}-z_{6,l+1}\mathcal{A}
\pa_{\mathcal{A}}\Gamma_{0|22}(V)+z_{22,l+1}\Gamma_{0|22}(V), \\
\label{5.2.2.15a}
&&s_{l+1}^{(2)}=\sum_{k=1}^5z_{k,l+1}\Gamma_{0|k}, \; \dot{z}_{22,l+1}=
\dot{z}_{23,l+1}=\dot{z}_{24,l+1}=\dot{z}_{25,l+1}=\dot{z}_{26,l+1}=0, \\
\label{5.2.2.16a}
&&z_{k_{dep},l+1}=\sum_a\big([X_{z|k_{dep},a}+Y_{z|k_{dep},a}\xi\pa_\xi]
z_{K_a,l+1}\big), \\
\label{e7a}
&&X_{z|k_{dep},a}=X_{q|k_{dep},a}, \; Y_{z|k_{dep},a}=Y_{q|k_{dep},a}.
\eeq
Due to (\ref{5.2.2.14}), (\ref{5.2.2.15}), (\ref{5.2.2.14a}) and (\ref{5.2.2.15a}) the
functional $\Gamma_{l+1,\mathrm{div}}$ (\ref{e6a}) is written in the form
\beq
\nonumber
&&\Gamma_{l+1,\mathrm{div}}=\sum_{k=6}^{21}(q_{k,l+1}+z_{k,l+1})\Gamma_{0|k}-
(q_{6,l+1}+z_{6,l+1})\mathcal{A}\pa_{\mathcal{A}}\Gamma_{0|22}(V)+ \\
\label{e6a2} &&\qquad\qquad
+(q_{22,l+1}+z_{22,l+1})\Gamma_{0|22}(V)+\chi\sum_{k=1}^5[(q_{k,l+1}+
z_{k,l+1})\Gamma_{0|k}]. \eeq Now, let us take $z_{k_{ind},l+1}$ as
\beq \label{e8} z_{k_{ind},l+1}=-q_{k_{ind},l+1}. \eeq Then, due to
(\ref{5.2.2.16a}), (\ref{e7}) and (\ref{5.2.2.16}), we obtain that
\beq \label{e9} z_{k_{dep},l+1}=-q_{k_{dep},l+1},
\eeq
and as a
consequence
\beq
\Gamma_{l+1,\mathrm{div}}=0.
\eeq

Thus we have found the parameters $Z_{k_{ind}}$,
\beq
\label{ca1a}
Z_{k_{ind}}=Z_{k_{ind}}^{[l+1]}+O(\eta^{l+2}), \quad
Z_{k_{ind}}^{[l+1]}= Z_{k_{ind}}^{[l]}+z_{k_{ind},l+1}, \quad
\dot{Z}_{k_{con}}^{[l+1]}=0
\eeq
 so that the $(l+1)$-loop
approximation for $\Gamma$,
$\Gamma^{[l+1]}=\sum_{n=0}^{l+1}\eta^n\Gamma_n$, is a finite
functional. Note that the parameters $z_{k_{ind},l+1}$ and
$z_{k_{dep},l+1}$ given by eqs. (\ref{e8}) and  (\ref{5.2.2.16a}),
respectively, are defined unambiguously by the divergences.

It is evident that this method works for any $l$, in particular for
$l=0$, so that by using the loop induction method in Feynman
diagrams for the functional $\Gamma$, we arrive at the following
statement: for the $l$-loop approximation $\Gamma^{[l]}$ of the
functional  $\Gamma$ defined by the relations (\ref{b1}), (\ref{b2})
with  arbitrary positive integer $l$ ,
\beq
\Gamma^{[l]}=\sum_{n=0}^l\eta^n\Gamma_n,
\eeq
there exist the
uniquely  defined parameters $Z_{k_{ind}}^{[l]}$, and parameters
$z_{k_{dep}}^{[l]}$ defining by eq. (\ref{5.2.2.16a}),
\beq
\label{e12} \dot{Z}_{k_{con}}^{[l]}=0,
\eeq
such that the functional
$\Gamma^{[l]}$ does not contain divergences and $\Gamma$ satisfies
the equations (\ref{b4}) - (\ref{b7}).

\section{Gauge independence of physical parameters}
\noindent
In that section, we find relations between some parameters
of the action $S_R$ and the standard renormalization constants.
Within  the expression for $S_R$, we restrict ourselves only by
desired vertexes in symbolic notation
\beq
 \nonumber
&&S_R=\int
dx\left(Z_{22}Z_6^{-2}\pa A\pa A+gZ_{22}Z_6^{-3}A^2\pa A+
Z_{13}\overline{\psi}\pa\psi+mZ_{15}\overline{\psi}\psi+Z_{16}\pa\varphi
\pa\varphi+ \right.\\
\label{s1} &&\qquad\qquad\qquad  +M^2Z_{19}\varphi^2+\lambda
Z_{20}\varphi^4\left.+\vartheta Z_{21}\varphi
\overline{\psi}\psi...\right),
 \eeq
 where the ellipsis means the
rest vertexes. As the propagators of the fields
 $A$, $\psi$ and $\varphi$ are finite, they should be considered as
 renormalized fields. Then, we find:
\beq
Z_A=Z_{22}^{1/2}Z_6^{-1}, \quad Z_\psi=Z_{13}^{1/2},\quad Z_{\varphi}=Z_{16}^{1/2},
\eeq
where $Z_A$, $Z_\psi$ and $Z_\varphi$ are the renormalization constants of the
bare fields $A_0$, $\psi_0$ and $\varphi_0$. The coefficient of the second
vertex in the expression (\ref{s1}) gives the renormalization for vertex
$gA^2\pa A$,
\beq
Z_{gA^2\pa A}=Z_{22}Z_6^{-3}.
\eeq
Analogously,
\beq
Z_{m\overline{\psi}\psi}=Z_{15}, \; Z_{M^2\varphi^2}=Z_{19}, \;
Z_{\lambda\varphi^4}=Z_{20}, \; Z_{\vartheta\varphi\overline{\psi}\psi}=Z_{21}.
\eeq

Then we find
\beq
&&g_0=Z_gg, \quad Z_g=Z_{gA^2\pa A}Z_A^{-3}=Z_{22}^{-1/2}, \\
&&m_0=Z_mm=Z_{m\overline{\psi}\psi}Z_{13}^{-1}=Z_{23}, \\
&&M_0^2 =Z_{M^2}M^2, \quad  Z_{M^2}=Z_{19}Z_{16}^{-1}=Z_{24}, \\
&&\lambda_0= Z_{\lambda}\lambda, \quad  Z_\lambda=Z_{20}Z_\varphi^{-4}=Z_{25}, \\
&&\vartheta_0=Z_\vartheta\vartheta, \quad  Z_\vartheta=
Z_{21}Z_\varphi^{-1/2}Z_\psi^{-1}=Z_{26}.
\eeq

It follows from the equations (\ref{e12}) that the renormalization constants
of physical parameters  $g$, $m$, $M^2$, $\lambda$ and $\vartheta$ do not
depend on gauge,
\beq
\pa_\xi Z_g=\pa_\xi Z_m=\pa_\xi Z_{M^2}=\pa_\xi Z_\lambda=
\pa_\xi Z_\vartheta=0.
\eeq

\section{Summary}
\noindent

In the present paper, within the background field formalism
\cite{DeW,AFS,Abbott}, we have studied the renormalization and the
gauge dependence of the $SU(2)$ Yang-Mills theory with the
multiplets of massive spinor and scalar fields. The corresponding
master-action of the BV-formalism \cite{BV,BV1} has been extended
with the help of additional fermion vector field $\theta$ and
fermion constant parameter $\chi$. The action introduced is
invariant under global supersymmetry and gauge transformations
caused by the background vector field ${\cal B}$ appearing in the
background field formalism. These symmetries allowed one to reduce,
at the quantum level, the analysis of the renormalization and the
gauge dependence problem for solutions to the extended
master-equation and the gauge dependence equation.  In comparison
with our previous investigations of the multiplicative
renormalization of the Yang-Mills theories \cite{BLT-YM}, recent
study involves the scalar fields which can be responsible for
generating masses to physical particles through the mechanism of
spontaneous symmetry breaking \cite{Higgs}.

The proofs of multiplicative renormalizability and gauge
independence of renormalization constants are based on the
possibility to expand the extended effective action in loops, as
well as to use the minimal subtraction scheme as to eliminate
divergences. In addition, we propose the existence of a
regularization preserving the used symmetries. Among the results
obtained, we emphasize the rigorous proof of the gauge independence
of all the physical parameters of the theory under consideration, to
any order of loop expansions.

\section*{Acknowledgments}
\noindent
We would like to thank Klaus Bering of Masaryk University for useful discussions and kindly help with
References. Also we would like to thank  anonymous Referee B who paid  our attention
to the paper \cite{PS}.
P.M. Lavrov  is grateful to the Physics Department  of the Federal University of Juiz de Fora (MG,
Brazil) for warm hospitality. The work of I.A. Batalin and I.V. Tyutin is
supported in part by the RFBR grant 17-02-00317. The work of P.M.
Lavrov is supported partially by the Ministry of Education and Science of
the Russian Federation, grant  3.1386.2017 and by the RFBR grant
18-02-00153.
\\

\end{document}